\documentclass[aps,a4paper,prd,amssymb,amsmath,eqsecnum,showpacs,twocolumn] {revtex4}
\usepackage{color}
\usepackage{ulem}
\usepackage{scalerel}
\usepackage{tikz}
\usetikzlibrary{svg.path}

\definecolor{orcidlogocol}{HTML}{A6CE39}
\tikzset{
  orcidlogo/.pic={
    \fill[orcidlogocol] svg{M256,128c0,70.7-57.3,128-128,128C57.3,256,0,198.7,0,128C0,57.3,57.3,0,128,0C198.7,0,256,57.3,256,128z};
    \fill[white] svg{M86.3,186.2H70.9V79.1h15.4v48.4V186.2z}
                 svg{M108.9,79.1h41.6c39.6,0,57,28.3,57,53.6c0,27.5-21.5,53.6-56.8,53.6h-41.8V79.1z M124.3,172.4h24.5c34.9,0,42.9-26.5,42.9-39.7c0-21.5-13.7-39.7-43.7-39.7h-23.7V172.4z}
                 svg{M88.7,56.8c0,5.5-4.5,10.1-10.1,10.1c-5.6,0-10.1-4.6-10.1-10.1c0-5.6,4.5-10.1,10.1-10.1C84.2,46.7,88.7,51.3,88.7,56.8z};
  }
}

\newcommand\orcidicon[1]{\href{https://orcid.org/#1}{\mbox{\scalerel*{
\begin{tikzpicture}[yscale=-1,transform shape]
\pic{orcidlogo};
\end{tikzpicture}
}{|}}}}
\usepackage{hyperref}

\begin{document}
\author{Thomas M\"adler\orcidicon{0000-0001-5076-3362}}, \email{thomas.maedler_.at._mail.udp.cl}

\newcommand{\red}[1]{\textcolor{red}{#1}}
\newcommand{\blue}[1]{\textcolor{blue}{#1}}

\affiliation{Escuela de Obras Civiles and
Instituto de Estudios Astrof\'isicos,
Facultad de Ingenier\'{i}a y Ciencias, Universidad Diego Portales, Avenida Ej\'{e}rcito
Libertador 441, Casilla 298-V, Santiago, Chile.
}
\author{Emanuel Gallo\orcidicon{0000-0002-8974-5134}}
\email{egallo_.at._unc.edu.ar}

\affiliation{FaMAF, UNC; Instituto de Física Enrique Gaviola (IFEG), CONICET, \\
Ciudad Universitaria, (5000) C\'ordoba, Argentina. }

\title{Slowly rotating Kerr metric derived from the Einstein equations in affine-null coordinates}

\begin{abstract}
Using a quasi-spherical approximation of an affine-null metric adapted to an asymptotic Bondi inertial frame, we present high order approximations of the metric functions in terms of the specific angular momentum for a slowly rotating stationary and axi-symmetric vacuum spacetime. 
The metric is obtained by following the procedure of integrating the hierarchy of  Einstein equations in a characteristic formulation utilizing master functions for the perturbations. It is further verified its equivalence with the Kerr metric in the slowly rotation approximation by carrying out an explicit transformation between the Boyer-Lindquist coordinates to the employed affine-null coordinates. 
A peculiar feature of the derivation is that in the solution of the perturbation equations for  every order  a new integration constant appears which cannot be set to zero using asymptotical flatness or regularity arguments. However these additional integration constants can be absorbed into the overall Komar mass and Komar angular momentum of a slowly rotating black hole.  
\end{abstract}

%\pacs{ 04.20.-q, 04.20.Cv, 04.20.Ex, 04.25.D- }

\maketitle
\section{Introduction} 
At the dawn of the 'Golden Era of General Relativity' in the 60ties of the last century, two important spacetime metrics were found, the Bondi-Sachs metric \cite{1960Natur.186..535B,1962RSPSA.269...21B,1962RSPSA.270..103S} and the Kerr metric \cite{1963PhRvL..11..237K,1965JMP.....6..918N}. The first settled the question that an isolated system looses mass via gravitational radiation and that this effect is a non-linear effect of General Relativity; while the second describes  a stationary and rotating  isolated black hole that is expected to be the end product of a gravitational collapse of a massive star or a merger of two compact objects. 

One of the defining features of the Bondi-Sachs metric is that  one  coordinate is constant along a family of null hypersurfaces while a radial coordinate along these null hypersurfaces is an areal distance that can be related to a luminosity distance \cite{2013GReGr..45.2691J}. 
Indeed, the first long term stable evolution of black hole space times were made using such families of null hypersurfaces in a null cone-world tube formalism \cite{1998PhRvD..57.4778G}, also see \cite{2012LRR....15....2W,2016SchpJ..1133528M} for review. 
Apart from usage in numerical relativity simulations, the Bondi-Sachs metric is now frequently used in high energy physics addressing questions of the AdS/CFT correspondence \cite{2010JHEP...05..062B} (and citations thereof). It also became popular to discuss gravitational wave memory effects \cite{2016JHEP...12..053P,2016CQGra..33q5006M,2017PhRvD..95h4048N,2018CQGra..35c5009M,2019CQGra..36i5009M}.
A pleasant property of the Bondi-Sachs formalism is that  the Einstein equations can be solved in a hierarchical manner when initial data on a null hypersurface and boundary conditions at a null hypersurface \cite{2002PhRvD..65h4016P}, world tube \cite{1996PhRvD..54.6153B} or vertex\cite{1994JMP....35.4184G,2002PhRvD..65f4038S,2013CQGra..30e5019M} are given. However, the  radial coordinate of the Bondi-Sachs metric has the unpleasant property that it breaks down when an apparent horizon forms due to the focusing of the surface-forming null rays and their vanishing expansion. This can be overcome in choosing an affine parameter as radial coordinate, because an affine parameter only becomes singular at a caustic.  
But, the Einstein equations resulting from an affine-null metric do not provide the hierarchical structure as the Bondi Sachs metric \cite{2016SchpJ..1133528M} and the hierarchical structure needs to be reestablished by various new definitions of variables \cite{2013PhRvD..87l4027W,2019PhRvD..99j4048M,2021PhRvD.104h4048G}. 
Moreover, it turns out that also the  hierarchy of equations in the affine-null metric formulation breaks down in the events of apparent horizon formation, but fortunately the equations can be regularized so that it is possible to follow up the formation of black holes up to singularity \cite{2019PhRvD.100j4017C,Baake2023}.

 Despite the success and popularity of the Bondi-Sachs  metric in the various areas, an explicit  closed analytical  representation of the Kerr metric in Bondi-Sachs form without bad behaviour in the exterior region or related metrics with one or two null coordinates is missing. Various attempts have been made to derive a null metric representation, numerically \cite{2006PhRvD..73h4023B,2021PhRvD.104b4049A} as well as analytically \cite{2003CQGra..20.4153F, 2021arXiv210801098J,Hayward:2004ih}. 
In {\it all} of the approaches, the authors start out with the Kerr metric and then calculate the respective null metric via a coordinate transformation. 
After these transformations the resulting metric can still posses a conical singularity at the axis of symmetry  (see \cite{2021PhRvD.104b4049A} for a complete discussion). In addition, the final metric is determined by integrals of non-elementary functions. 
Arga\~naraz and Moreschi's approach \cite{2021PhRvD.104b4049A} differs to the aforementioned ones that the authors aim to find a double--null representation of the Kerr metric by geometrically adopting the coordinates to in- and outgoing null geodesics adapted to the center of mass \cite{Gallo:2014jda}. In this way, the authors were successful in finding  null coordinates that are not only regular at every point of the external communication region (unlike the previous formulations) but also that they are regular at the event horizon, thus allowing a way to study the evolution of different matter fields (as scalar fields) in such background even when they cross the event horizon\cite{Arganaraz:2022mks}.
Unfortunately, even in their construction  arises a differential equation that needs to be solved numerically and an explicit closed form representation of the double null version of the Kerr metric is not possible. The work of Bai and collaborators  \cite{2007PhRvD..75d4003B,2007PhRvD..76j7501G} also starts with the Kerr metric (in Boyer-Lindquist coordinates) and then makes coordinate transformation to a Bondi-Sachs metric valid near future null infinity (in a compactified version of the metric). The authors are able to calculate the Newman-Penrose quantities and multipoles  at large distances and show the peeling property of the Weyl tensor at large radii and the vanishing of the so-called Newman-Penrose constants. 

In this article, {\it in  contrast} to all the previous works which start with the Kerr metric expressed in Boyer-Lindquist coordinates and attempt to find a null coordinate version of it, we will {\it directly solve} the   Einstein equations in a characteristic formulation based on an affine-null metric formulation of the Einstein equations. In addition, inspired by the Hartle-Thorne methods for obtaining solutions for slowly rotating compact stars \cite{1967ApJ...150.1005H}, we will employ a quasi-spherical approximation of the field equations to find a high order approximation of the Kerr metric in out-going  polar null coordinates. 
To obtain our solution, we assume stationarity and axial symmetry. 
We further require an asymptotic inertial observer as well as that that Weyl scalar $\Psi_0$ is regular everywhere where the background solution is regular.
A study of vacuum stationary metrics with a smooth future null infinity in affine-null coordinates has recently be carried out by Tafel in \cite{Tafel:2021kza} by considering power series of the metric components in terms of the inverse affine distance.

Throughout the article, we will use signature $+2$, units $G=c=1$ and the Einstein sum convention for indices as well as products of associated Legendre polynomials.

The article is organised as follows:  Sec.~\ref{sec:aff_null} recalls the affine-null metric formulation, makes the necessary symmetry assumptions for archiving our goal and defines the perturbative variables; in Sec.~\ref{sec:sol_perturbation}, we determine the background model (Sec.~\ref{sec:back}), define useful recursively re-appearing functions in the perturbation analysis (Sec.~\ref{sec:operators}), solve the perturbation equations (Sec.~\ref{sec:sol_lin_pert}-\ref{sec:sol_quartic}) and in Sec.~\ref{sec:combine_Komar} the affine-null metric functions for the null are expressed in terms of the mass and specific angular momentum, in Sec.~\ref{sec:affine_null_kerr_trafo}, to verify our results, we calculate the affine-null version of Kerr metric in a Bondi frame via a coordinate transformation with a method adopted from \cite{2007PhRvD..75d4003B},
in Sec.~\ref{sec:affine_null_kerr_trafo_hor} the position of the outer ergosphere and (past) event  horizon of the black hole are discussed and   Sec.~\ref{sec:conclusion} contains the final discussion of our work.
The article finishes with two appendices: App.~\ref{sec:app_LP}  lists  relations between associated Legendre polynomials 
and 
App.~\ref{sec:app_komar_charge} presents a derivation of the expression of the Komar charges relevant for this work.
\section{Affine-null metric formulation for stationary and  axial symmetric spacetimes }\label{sec:aff_null}
Here we review the necessary properties of characteristic initial value formulation of the Einstein equations in affine-null coordinates, discuss the implications of the imposed  symmetry assumptions and  present the notation used in our analysis.

Taking coordinates $x^a = (u,\lambda, x^A)$, where $u$ is 
an out--going null coordinate,  
$\lambda$ an affine parameter, and $x^A$ are angular coordinates, a generic line element for an affine-null metric defined with respect to a family of outgoing null hypersurfaces $u=const$ is \cite{2001PhRvD..64b4010G,2013PhRvD..87l4027W,2019PhRvD..99j4048M,2021PhRvD.104h4048G}
\begin{eqnarray}
   \lefteqn{ g_{ab}dx^adx^b 
   =  -W du^2
   -2 du d\lambda 
   }
    \nonumber\\
    && 
    \quad+ R^2h_{AB}(dx^A - W^Adu)(dx^B - W^Bdu). 
    \label{eq:an_metric}
\end{eqnarray}
The determinant  $\det(h_{AB}) = \det(q_{AB}) =\sin^ 2\theta$ is the determinant of a round unit sphere metric $q_{AB}$. We remark that the affine parameter $\lambda$ is chosen along the outgoing null hypersurfaces $u=const$ such that $\nabla^a u\nabla_a \lambda = -1$ everywhere along the rays generating the  null hypersurfaces  $u=const$ \cite{2001PhRvD..64b4010G}. Consequently $h_{AB}$ is transverse-traceless and has only two degrees of freedom. 
Thus, the function $R$ relates  to the area of cuts $du=d\lambda=0$.
The non-zero components of the  inverse metric  are given by  
\begin{equation}\label{eq:contrav_metric}
    g^{u\lambda} = -1\;\;,\;\;
    g^{\lambda\lambda} = W\;\;,\;\;
    g^{\lambda A} = -W^A\;\;,\;\;
    g^{AB} = \frac{h^{AB}}{R^2},
\end{equation}
where $W^A=(W^\theta, W^\phi)$ and $h_{AB}h^{BC} = \delta^{C}_{A}$ and 
in particular \cite{1966RSPSA.294..112V}
\begin{eqnarray}
h_{AB}dx^Adx^B&=&
 \Big(e^{2\gamma} d\theta^2 +\frac{\sin^ 2\theta}{ e^{2\gamma}}d\phi^ 2 \Big)\cosh(2\delta)\nonumber\\
 &&
 +2\sin\theta\sinh(2\delta)d\theta d\phi\;\;.
 \label{eq:vdB}
\end{eqnarray}
A complex null dyad to represent the 2-metric $h_{AB}$ like $h_{AB} = m_{(A}\bar m_{B)}$ with $m^Am^Bh_{AB} = m^A\bar m^Bh_{AB}-1=0 $
is
\begin{align}
m^A\partial_A = &\frac{1}{\sqrt{2}e^{\gamma}}\Big(\cosh\delta - i\sinh\delta\Big){\partial_\theta}\nonumber
\\ &+ \frac{i e^{\gamma}}{\sqrt{2}\sin\theta}\Big(\cosh\delta + i\sinh\delta\Big)\partial_\phi,
\end{align}
Like in any Bondi-Sachs type metric \cite{2016SchpJ..1133528M}, the vacuum field equations  $R_{ab} = 0$ with $R_{ab}$ being the Ricci tensor can be grouped into supplementary equations $S_i=0$ with
\begin{equation}\label{supplentary_notation}
    S_i = (R_{uu}, R_{u\theta}, R_{u\phi}),
\end{equation}
 one trivial equation, $R_{u\lambda}=0$, and the six main equations 
     $H^{(\gamma)}_{K} = 0 $, $K\in(1,2,3,4))$ and $H^{(\delta)}_k=0$, $k\in(1,2))$
 with  
 \begin{equation}\label{main_notation}
     \begin{split}
     H^{(\gamma)}_{K} =&  \left(R_{\lambda\lambda}, R_{\lambda \theta},h^{AB}R_{AB}, \Re e (m^{A}m^{B}R_{AB})\right ) ,   \\
     H^{(\delta)}_{k} =&  \left(R_{\lambda \phi},  \Im m(m^{A}m^{B}R_{AB}) \right) ,   \\
     \end{split}    
 \end{equation}
with $\Re e(x)$ and $\Im m(x)$ the real an imaginary part of $x$ respectively.
We assume that the spacetime is axisymmetric and stationary with  associated Killing vectors fields  $\partial_u$ and $\partial_\phi$. Therefore the metric functions do not depend on $u$ and $\phi$. The Killing symmetries imply two conserved quantities, the Komar mass, $K_m$,  and the Komar  angular momentum, $K_L$, which can be calculated from their respective  integrals (also see App.~\ref{sec:app_komar_charge})
\begin{equation}\label{eq:komm} 
K_m:=K(\partial_u) = \frac{1 }{8\pi}\lim_{\lambda\rightarrow\infty}\oint \Big(   
        W_{,\lambda}
      -R^2h_{AB}W^AW^B_{,\lambda}
      \Big) R^2 d^2q  
\end{equation}
while  for the  axial Killing vector we have
\begin{equation}\label{eq:Lang}
K_L:=K(\partial_\phi)
=
-\frac{1}{16\pi}\lim_{\lambda\rightarrow\infty}\oint \Big(R^4 h_{\phi B}W^B_{,\lambda}\Big) d^2q 
\end{equation}
where $dq = \sin\theta d\theta d\phi$ is the surface area element of the unit sphere.

Let us assume there is a smooth one parameter family of stationary and axially symmetric metrics $g_{ab}(\varepsilon)$, where $\varepsilon$ is a small parameter such that $\varepsilon=0$  is a corresponds to a (static) spherically symmetric spacetime solution of the vacuum Einstein equations. Then there is an  expansion of the metric fields like
\begin{widetext}
\begin{subequations}\label{expansions_fields}
\begin{eqnarray}
  R(\lambda, \theta) &=& r(\lambda) + R_{[1]}(\lambda,\theta) \varepsilon+R_{[2]}(\lambda, \theta) \varepsilon^2
  +R_{[3]}(\lambda, \theta) \varepsilon^3+O(\varepsilon^4),\label{eq:Rgen}\\
W(\lambda, \theta) &=& V(\lambda)
+  W_{[1]}(\lambda,\theta)\varepsilon
+ W_{[2]}(\lambda, \theta)\varepsilon^2
+ W_{[3]}(\lambda, \theta)\varepsilon^3 
+O(\varepsilon^4), \\
 W^A(\lambda, \theta) &=&      
   W^A_{[1]}(\lambda,\theta)\varepsilon
 + W^A_{[2]}(\lambda, \theta)\varepsilon^2 
 + W_{[3]}^A(\lambda, \theta)\varepsilon^3
+O(\varepsilon^4),\\
 \gamma(\lambda, \theta)&=&\gamma_{[1]}(\lambda, \theta)\varepsilon + \gamma_{[2]}(\lambda, \theta)\varepsilon^2
 + \gamma_{[3]}(\lambda, \theta)\varepsilon^3
  + O(\varepsilon^4), \label{eq:gammd}\\
 \delta(\lambda, \theta)&=&  
 \delta_{[1]}(\lambda, \theta)\varepsilon+
 \delta_{[2]}(\lambda, \theta)\varepsilon^2
 + \delta_{[3]}(\lambda, \theta)\varepsilon^3
+ O(\varepsilon^4).\label{eq:deltad}
\end{eqnarray}
\end{subequations}

\end{widetext}
 Inserting \eqref{expansions_fields} in \eqref{eq:komm}  and \eqref{eq:Lang} implies $K_m = O(\varepsilon^0)$ and $K_L = O(\varepsilon) $. We make the requirements 
 \begin{equation}
     K_m(\varepsilon) = K_m(-\varepsilon) \;\;,\;\;
     K_L(\varepsilon) = -K_L(-\varepsilon).\label{eq:lab}
 \end{equation}
These conditions imply that under the change  ${\varepsilon\rightarrow -\varepsilon}$ the sense of rotation is reversed (recall that ${K(\partial_\phi) = -K(\partial_{(-\phi)})}$).
 From the metric \eqref{eq:an_metric}, we see that the 2-surfaces with $u=u_0$ and $\lambda=\lambda_0$, defined such that $R(u_0,\lambda_0,\theta )=$const have the induced metric $R^2h_{AB}dx^Adx^B$ with area $4\pi R^2(u_0,\lambda_0)$.
 We assume that the area of these 2-surfaces is invariant under the change $\varepsilon\rightarrow -\varepsilon$, which implies that  $R^2$ is an even function of $\varepsilon$. Therefore $R$ is either an even or an odd function of $\varepsilon$. However, if $R$ were an odd function, we had  $R(\varepsilon=0)=0$, which is a non admissible solution.
 In addition, $ds^2(\partial_\phi,\partial_\phi)$ and $ds^2(\partial_\theta,\partial_\theta)$ must be independent of the sense of rotation implying that $h_{\phi\phi}$ and $h_{\theta\theta}$ are even. 
 However, due to the  frame dragging effect $ds^2(\partial_\theta,\partial_\phi)$ must depend on the sense of rotation. Therefore $h_{\theta\phi}$ is an odd function of $\varepsilon$. 
 Using similar arguments,  because the Komar angular momentum $K_L$ is an odd function of $\varepsilon$  and taking into account \eqref{eq:Lang} and the parity behaviour of $h_{AB}$ and $R^2$, we have that $W^\theta$ is even and $W^\phi$ odd. Similarly, since  $K_m$ must be a even function of $\varepsilon$, $W$  must be even in $\varepsilon$. Therefore,
 \begin{subequations}\label{eq:sym_metric}
    \begin{align}
     R_{[2n+1]} &=W_{[2n+1]}= 0,\\
     W^\theta_{[2n+1]}&= 0,\\
     {W^\phi_{[2n]}}&=0,\\
     \gamma_{[2n+1]}&=\delta_{[2n]}=0.\label{eq:gd2n1}
 \end{align}
 \end{subequations}
 
 To arrive at the last conditions \eqref{eq:gd2n1} we have taken into account the odd parity of   $h_{\theta\phi}$, which gives us  $\sinh(\delta(\varepsilon))=-\sinh(\delta(-\varepsilon))$. Hence,  $\delta$  must be odd in  $\varepsilon$. Similarly, for  $h_{\theta\theta}$ and $h_{\phi\phi}$ be even, $\gamma(\varepsilon)$ must satisfies $e^{2\gamma(\varepsilon)}=e^{2\gamma(-\varepsilon)}$, which implies that $\gamma$ is a even function of $\varepsilon$.
 
 We conclude 
 \begin{subequations}\label{pert_no_sym}
\begin{eqnarray}
  R &=& r +R_{[2]} \varepsilon^2 +R_{[4]} \varepsilon^4+O(\varepsilon^6),\\
W  &=& V + W_{[2]} \varepsilon^2 + W_{[4]} \varepsilon^4 +O(\varepsilon^6),\\
 W^\theta  &=&   W^\theta_{[2]} \varepsilon^2 +W^\theta_{[4]} \varepsilon^4+O(\varepsilon^4),\\
 W^\phi  &=&  W^\phi_{[1]} \varepsilon+W_{[3]}^\phi \varepsilon^3+O(
 \varepsilon^5),\\
 \gamma&=&  \gamma_{[2]} \varepsilon^2
 + \gamma_{[4]} \varepsilon^4+ O(\varepsilon^6),\\
 \delta&=&\delta_{[1]}\varepsilon+
 \delta_{[3]}\varepsilon^3+ O(\varepsilon^5).
\end{eqnarray}
\end{subequations}
A similar expansion was made by Hartle \cite{1967ApJ...150.1005H} in the derivation of a metric for  slowly rotating stars using a a 3+1 decomposition of the metric.
From  \eqref{expansions_fields}  follows that the Ricci tensor has the expansions
\begin{align}    R_{ab} = R_{[0]ab} + R_{[1]ab}\varepsilon+
    R_{[2]ab}\varepsilon^2+
    R_{[3]ab}\varepsilon^3+ ...
\end{align}
In fact, with the notation
$
    {f_{[i]}\in\{\gamma_{[i]}, \delta_{[i]},
    R_{[i]},
    W^A_{[i]},
    W_{[i]}\}}
$,  
it turns out  for a perturbation at order  $n>1$ that
\begin{align}
    S_{[n]i} =& \hat{S}_{i} (f_{[n]}) + s_{[i]}(f_{[m<n]})\label{eq:principal_S}\\
    H^{(\gamma)}_{K}=& \hat{H}^{(\gamma)}_{K} (f_{[n]}) + h^{(\gamma)}_{K}(f_{[m<n]})\label{eq:principal_hg}\\
    H^{(\delta)}_{k}=& \hat{H}^{(\delta)}_{k} (f_{[n]}) + h^{(\delta)}_{k}(f_{[m<n]})\label{eq:principal_hd}
 \end{align}
where $\hat{S}_{i} $, $\hat{H}^{(\gamma)}_{K}$ and  $\hat{H}^{(\delta)}_{k}$ are linear differential operators of the indicated arguments.
The functions $s_{[i]}$, $h^{(\gamma)}_{K}$ and $h^{(\delta)}_{k}$  are nonlinear functions of the lower order perturbations $f_{[m]}$ for $m<n$.

%\red{structure}

For the  computations, it is useful to change the angular coordinate according to $y=-\cos\theta$, introduce ${s(y) = \sqrt{1-y^2}}$ and transform $W^\theta=s^{-1}W^y$. In addition, for a perturbation at order $n$ it will shown useful to make the following decomposition of 
the perturbation $f_{[n]}$ in terms of associated Legendre polynomials, $P^m_\ell(y)$,
\begin{align}
    R_{[n]}(\lambda,y) =&  R_{[n.\ell]}(\lambda)P^0_\ell(y)\\
    W^y_{[n]}(\lambda,y) =&  W^\phi_{[n.\ell]}(\lambda)\left[s(y) P^1_\ell(y)\right]\\
    W^\phi_{[n]}(\lambda,y) =&  W^\phi_{[n.\ell]}(\lambda)\left[\frac{ P^1_\ell(y)}{s(y)}\right]\\
    W_{[n]}(\lambda,y) =&  W_{[n,\ell]}(\lambda)P^0_\ell(y)\\
    \gamma_{[n]}(\lambda,y) =&  \gamma_{[n.\ell]}(\lambda)P^2_\ell(y)\\
    \delta_{[n]}(\lambda,y) =&  \delta_{[n.\ell]}(\lambda)P^2_\ell(y)\;\;, 
\end{align}
in which we also apply the Einstein sum convention over $\ell$, the respective harmonics  of the associated Legendre polynomials.
We remark that this decomposition with respect to the associated Legendre polynomials is in fact a decomposition in terms of axi-symmetric spin-weighed harmonics (up to normalisation) obtained by setting $m=0$ in the standard $_sY_{\ell m}(y, \phi)$.

\section{Solution of the Background and perturbation equations }\label{sec:sol_perturbation}
\subsection{Solution background equations}\label{sec:back}
The main equations for the background model are
\begin{subequations}
\begin{align}
    0&=\frac{r_{,\lambda\lambda}}{r}\\
    0&=[(r^2)_{,\lambda} V - 2\lambda]_{,\lambda}\;\;,\label{eqn_backgroundV}
\end{align}
\end{subequations}
From which we deduce  

\begin{eqnarray}
r(\lambda)&=&r_1\lambda  + r_0
\end {eqnarray}
where $r_1$ and $r_0$ are integration constants, however since we have the freedom of rescaling the affine parameter $\lambda\rightarrow \alpha\lambda + \beta$, so that we can take without loss of generality 
\begin{subequations}\label{background_sol}
    \begin{equation}
    r=\lambda.
\end{equation}
Next integration of \eqref{eqn_backgroundV} yields
\begin{eqnarray}
%r(\lambda)&=&\lambda\;\;,\;\;
%\\
V(\lambda)=   1-\frac{A}{2\lambda}.
\end {eqnarray}
\end{subequations}
with $A$ an integration constant. 

 The resulting spacetime is the Schwarzschild metric  in outgoing Eddington-Finkelstein coordinates, with a total Bondi mass $m_0$ related to the integration constant $A$ by $A=4m_0$.
Moreover, $\lambda = A/2$ corresponds to location the past event horizon of the Schwarzschild horizon.

\subsection{Recurrent operators in the equations of the perturbations}
\label{sec:operators}

The principal part $\hat S_{i}(f_n)$ of the  supplementary equations in  \eqref{eq:principal_S} while recalling the notation $s^2(y) =1-y^2$ are
\begin{widetext}
\begin{subequations}
    \begin{eqnarray}
        \hat S_{1}(R _{[n]}, W_{[n]}, W^y_{[n]})&=&
        \frac{1}{2\lambda^2}  \left(1-\frac{A}{2\lambda}\right)\left(
        \lambda^2W_{[n],\lambda } 
        + \frac{AR_{[n]}}{\lambda }\right)_{,\lambda }
    +\frac{(s^2 W_{[n],y})_{,y}}{2\lambda^2}
    -\frac{A}{4\lambda^2}W^y_{[n], y}
    \nonumber\\
    &&
    \label{supp_S1_lin}\\
        \hat S_{2}(  W_{[n]}, W^y_{[n]})&=&
        \frac{1}{2\lambda^2}  \left(1-\frac{A}{2\lambda}\right)\frac{(\lambda^4 W^y_{[n]})_{,\lambda } }{s}
     +\frac{s}{2}W_{[n],\lambda y}
     +\frac{W^y_{[n]}}{s}
        \label{supp_S2_lin}\\
        \hat S_{ 3}( W^\phi_{[n]})&=&
         \frac{s^2}{2\lambda^2}  \left(1-\frac{A}{2\lambda}\right)(\lambda^4 W^\phi_{[n],\lambda })_{,\lambda }  
         +\frac{1}{2}\left(s^4W^\phi_{[n],y}\right)_{,y}\label{supp_S3_lin}
    \end{eqnarray}
\end{subequations}
Those for $\hat H^{(\gamma)}_{ K}$ in \eqref{eq:principal_hg} are
\begin{subequations}
    \begin{eqnarray}
        \hat H^{(\gamma)}_{1}(R _{[n]})&=&-\frac{2}{\lambda }R_{[n],\lambda\lambda}\\
        \hat H^{(\gamma)}_{2}(R_{[n]}, \gamma_{[n]}, W^y_{[n]})&=&
    \frac{1}{2\lambda^2}\frac{(\lambda^4 W^y_{[n],\lambda }) _{,\lambda }}{s}
    -s\left(\frac{R_{[n],y}}{\lambda }\right)_{,\lambda } 
    +\frac{(\gamma_{[n],\lambda }s^2)_{,y}}{s}
    \label{eq:hyp_E2_1}\\
        \hat H^{(\gamma)}_{3}(R_{[n]}, \gamma_{[n]}, W^y_{[n]}, W_{[n]})
        &=&
        - \left(\lambda  W_{[n]}\right)_{,\lambda }  
    - \left[\left(1 - \frac{A}{2\lambda}\right)(\lambda R_{[n]})_{,\lambda }\right]_{,\lambda }
    -\frac{\left(\lambda _{[n], y}s^2\right)_{,y}}{\lambda }
    +\frac{\left(\lambda^4  W^y_{[n]} \right)_{,\lambda y}}{2\lambda^2}
    \nonumber\\
    &&
   +  
   \frac{(\gamma_{[n],y}s^4)_{,y}}{s^2} -2\gamma_{[n]}
    \label{eq:hyp_E4_1}
    \nonumber\\
    &&
 \\
    \hat H^{(\gamma)}_{4} ( \gamma_{[n]}, W^y_{[n]})
    &=&
    - \left[\lambda\left(\lambda -\frac{A}{2}\right)\gamma_{[n],\lambda }\right]_{,\lambda }
    +\frac{s^2}{2} \left( \frac{\lambda^2W^y_{[n]}}{s^2}\right)_{,\lambda y}
    \label{eq:hyp_E5_1}
    \end{eqnarray}
\end{subequations}
and those for $\hat H^{(\delta)}_{k}$ of \eqref{eq:principal_hd}
\begin{subequations}
    \begin{eqnarray}
        \hat H^{(\delta)}_{1}( \delta_{[n]}, W^\phi_{[n]})&=& 
        \frac{s^2}{2\lambda^2}(\lambda^4 W^\phi_{[n],{\lambda }})_{,\lambda }
 +(\delta_{[n],\lambda }s^2)_{,y}
 \label{eq:hyp_E3_1}\\
        \hat H^{(\delta)}_{2}( \delta_{[n]}, W^\phi_{[n]})&=&
    - \left[\lambda\left(\lambda -\frac{A}{2}\right)\delta_{[n],\lambda }\right]_{,\lambda }
    -\frac{s^2}{2} \left( \lambda^2W^\phi_{[n]} \right)_{,\lambda y}
    \label{eq:hyp_E6_1}
    \end{eqnarray}
\end{subequations}
\end{widetext}
We observe that \eqref{eq:hyp_E2_1}  and \eqref{eq:hyp_E5_1} as well as \eqref{eq:hyp_E3_1} and \eqref{eq:hyp_E6_1} can be combined (see e.g. in \cite{2013PhRvD..87j4016M}) 
to two fourth order (master) equations
\begin{subequations}
\begin{align}
    0=&\mathcal{ M}(\gamma_{[n]}) -s^2 R_{[n],\lambda\lambda yy}  \label{eq:master_g1}\\
    0=&\mathcal{ M}(\delta_{[n]})\label{eq:master_d1}
\end{align}
\end{subequations}
where 
\begin{align}
  \mathcal{ M}(F):=&  \frac{1}{\lambda^2}[\lambda^4 (\lambda  F)_{,\lambda\lambda\lambda}]_{,\lambda }
    + [(\lambda F)_{,\lambda\lambda y}s^2]_{,y}
    \nonumber\\
   &
    + \left(\frac{A}{2\lambda}+2-\frac{4}{s^2}\right)(\lambda F)_{,\lambda\lambda}
    -\frac{A}{2}\left[\lambda (\lambda  F)_{,\lambda\lambda\lambda}\right]_{,\lambda }
\end{align}
We emphasize that Eqs. \eqref{eq:master_g1} and \eqref{eq:master_d1} (similarly to the Teukolsky master equations in 3+1 perturbation theory) are the key equations to solve the system, because they provide the initial data $\gamma_{[n]}$ or $\delta_{[n]}$ needed to integrate the hypersurface equations of the characteristic initial value problem. 

\subsection{First order perturbations}
\label{sec:sol_lin_pert}
Since $\gamma_{[1]}$, $R_{[1]}$, $W^y_{[1]}$ and $W_{[1]}$ are zero, we only have to consider the equations
\begin{align}
    0=&\hat S_{3}(\delta_{[1]}, W^\phi_{[1]})\label{supp_lin} \\
    0=&\hat H^{(\delta)}_{1}(\delta_{[1]}, W^\phi_{[1]})\label{hyp_Wphi_lin}\\    
    0=&\hat H^{(\delta)}_{2}(\delta_{[1]}, W^\phi_{[1]})\label{ev_lin}
\end{align}
whose explicit form can be read off from \eqref{supp_S3_lin}, \eqref{eq:hyp_E3_1} and \eqref{eq:hyp_E6_1}. The corresponding master equation is 
\begin{align}
 0=&  \frac{1}{\lambda^2}[\lambda^4 (\lambda  \delta_{[1]})_{,\lambda \lambda\lambda}]_{,\lambda }
    + [(\lambda \delta_{[1]})_{,\lambda \lambda y}s^2]_{,y}
    \nonumber\\
   &
    + \left(\frac{A}{2\lambda}+2-\frac{4}{s^2}\right)(\lambda \delta_{[1]})_{,\lambda \lambda}
    -\frac{A}{2}\left[\lambda (\lambda  \delta_{[1]})_{,\lambda \lambda\lambda}\right]_{,\lambda }
\end{align}
which is in fact a second order equation for the variable
\begin{equation}\label{def_psi}
    \psi_{[1]}:=(\lambda  \delta_{[1]})_{,\lambda \lambda}\;\;,
\end{equation}
namely
\begin{equation}\label{eq:nonLin_psi}
    \begin{split}
        0=& \lambda(2\lambda-A)\psi_{[1],\lambda \lambda} 
        + (8\lambda-A)\psi_{[1],\lambda } 
        +\frac{A\psi_{[1]}}{\lambda}\\
        &
        +2\left\{(s^2\psi_{[1],y})_{,y}
        +\left[\left(2-\frac{4}{s^2}\right)\right]\psi_{[1]}\right\},
    \end{split}
\end{equation}
 which admit a solution by separation of variables by setting $\psi_{[1]}(\lambda ,y) = p_{[1]}(\lambda ) S(y),$
\begin{align}
    0=&\lambda(2\lambda-A)p_{[1],\lambda \lambda} 
        + (8\lambda-A)p_{[1],\lambda } 
        +\left(\frac{A}{\lambda} 
        +2k\right)p_{[1]},
        \label{eq:g_ode}\\
    0=&\frac{d}{dy}\left[s^2\frac{dS}{dy}\right] 
    +\left(2-k-\frac{4}{s^2}\right) S,\label{eq:S_LEG}
\end{align}
with $k$ a constant.
Identifying  $2-k = \ell(\ell+1)$, we see that \eqref{eq:S_LEG} is an associated Legendre differential equation \eqref{aLP_ODE}, whose general solution is
\begin{align}
    S_{\ell}(y) = B_{0k}P(\ell, 2, y)+B_{1k}Q(\ell, 2, y),
\end{align}
where $P(\cdot)$ and $Q(\cdot)$ are the  Legendre functions of first kind and of second kind, respectively. 

Requiring a regular solution at the poles $y=\pm1$, imposes that $\ell$ must be a nonnegative integer and $B_{1k}=0$, because $P(\ell, 2, -1)$ blows up at the pole $y=-1$ and $Q(\ell, 2, \pm1)$ blows up at the poles $y=\pm1$.
Then the remaining Legendre function $P(\cdot)$ is the associated Legendre polynomial $P^2_\ell(y)$.  

To find a solution for   
\eqref{def_psi} and \eqref{eq:nonLin_psi}, we set 
\begin{equation}\label{eq:def_Psi_nonlin}
 \psi_{[1]}(\lambda ,y) = \psi_{[1.\ell]}(\lambda )P^2_\ell(y)\;\;,\;\;
 \delta_{[1]}(\lambda, y) = \delta_{[1.\ell]}(\lambda )P^2_\ell(y),   
\end{equation}
where a sum in $\ell$ is understood.
Note that $P^2_0(y)=P^2_1(y) = 0$ , consequently, $\delta_{[1.0]}=\delta_{[1.1]}=0$ %
without loss of generality.
Subsequent insertion into \eqref{eq:nonLin_psi} while using  eq.\eqref{aLP_ODE} 
gives us 
\begin{equation}\label{master_H1}
\begin{split}
    0 = &
    -\frac{1}{2}\lambda(A-2\lambda)\frac{d^2\psi_{[1.\ell]}}{d\lambda^2}
    +\left(4\lambda-\frac{A}{2}\right)
    \frac{d \psi_{[1.\ell]}}{d\lambda}\\
    &
    +\left[2-\ell(\ell+1)+\frac{A}{2\lambda}\right] \psi_{[1.\ell]}\;\;.
\end{split}
\end{equation}
Using the parameter transformation $x=\frac{4\lambda}{A}-1$, similar to \cite{1967ApJ...150.1005H}, we find
\begin{equation}
\begin{split}
    0=& (1-x)\frac{d^2 \psi_{[1.\ell]}}{dx^2}
    -\frac{4x + 2}{x + 1}\frac{d \psi_{[1.\ell]}}{dx}\\
    &+\frac{\ell(\ell+1)(x+1) - 2x-4}{(x + 1)^2} \psi_{[1.\ell]},
\end{split}
\end{equation}
which can also be written as
\begin{equation}\label{LegODE_x}
\begin{split}
    0=& \frac{d}{dy}\left[(1-x^2)\frac{d}{dx}(1-x) \psi_{[1.\ell]}\right]
    \\
    &+\left[\ell(\ell+1)-\frac{4}{1-x^2}\right](1-x)\psi_{[1.\ell]}.
\end{split}
\end{equation}
Eq. \eqref{LegODE_x} is an associated Legendre differential equation, like \eqref{aLP_ODE}, with the general solution
\begin{equation}\label{gen_sol_psi}
    \psi_{[1.\ell]}(x) = \frac{B_{[1.\ell]}P^2_\ell(x)+B_{[2.\ell]}Q^2_\ell(x)}{1-x}
\end{equation}
Inverting the parameter transformation from $x$ to  $\lambda$ yields the general solution of \eqref{master_H1} so that using \eqref{eq:def_Psi_nonlin}
\begin{equation}\label{gen_Sol_psi_LP}
\begin{split}
        \psi_{[1]}(\lambda ,y) = & \left(\frac{ AB_{[1.\ell]}
    }{2A-4\lambda}\right) P^2_\ell\left(\frac{4\lambda}{A}-1\right)P^2_\ell(y)\\
    &+ \left(\frac{ AB_{[2.\ell]}
    }{2A-4\lambda}\right) Q^2_\ell\left(\frac{4\lambda}{A}-1\right)P^2_\ell(y)
\end{split}
\end{equation}

The field $\psi_{[1]}$ is related to the Weyl scalar $\Psi_0$, 
\begin{equation}
    \Psi_0 = 
    -\frac{i\psi_{[1]}}{\lambda}\varepsilon 
    +O(\varepsilon^2)\;.
\end{equation}
Inspection of \eqref{gen_sol_psi} shows that $\Psi$ becomes infinite for $\lambda\rightarrow A/2$ and for $\lambda\rightarrow \infty$ if $\ell\ge2$. The first case corresponds to the unperturbed location of the horizon while the second one corresponds to the asymptotic region. Consequently, also $\Psi_0$ becomes infinite in these cases. We require regularity of the scalar curvature $\Psi_0$ at these locations,  which implies $B_{[1.\ell]}=B_{[2.\ell]}=0$. This leaves us with the trivial solution $\psi_{[1]}=0$. 

Integration of \eqref{def_psi} with this trivial solution while using \eqref{eq:def_Psi_nonlin} yields
\begin{equation}\label{sol_d1_ell}
\begin{split}
    \delta_{[1.\ell]}(\lambda )= B_{[0.\ell]}^\delta+\frac{B_{[1.\ell]}^\delta}{\lambda},
\end{split}
\end{equation}
where as aforementioned, since ${\delta_{[1.0]}=\delta_{[1.1]}=0}$ we get that $B_{[0.0]}^\delta=B_{[0.1]}^\delta=B_{[1.0]}^\delta=B_{[1.1]}^\delta=0$. 
These modes are physically irrelevant because $\delta_{[1]}$ is expressed by the angular base of $P^2_\ell-$associated Legendre polynomials and $P^2_0=P^2_1=0$. 
Since $\delta_{[1]}$ is now known, we are now in position to integrate the  hypersurface equation \eqref{hyp_Wphi_lin}. We insert \eqref{sol_d1_ell} into \eqref{hyp_Wphi_lin},
while using \eqref{rel1_aLP}, to find
\begin{equation}
     (\lambda ^4 W^\phi_{[1],{\lambda}})_{,\lambda }  = 
 2\lambda^2\left(\frac{d\delta_{[1.\ell]}}{d\lambda}\right)\frac{
    K_\ell P^1_\ell(y)}{s}\;\;.
\end{equation}
where
\begin{equation}
K_\ell =2-\ell(\ell+1) = (1-\ell)(2+\ell).
\end{equation}
Then setting 
\begin{align}
     W^\phi_{[1]}(\lambda , y) =& W^\phi_{[1.\ell]}(\lambda )\frac{P^1_\ell(y)}{s},
\end{align}
gives us
\begin{equation}
     \frac{d}{d\lambda}\left(\lambda ^4 \frac{W^\phi_{[1.\ell]}}{d\lambda}\right)   = 
   2B^\delta_{[1.\ell]}K_\ell ; 
\end{equation}
or after integration
\begin{equation}\label{sol_w31_ell}
\begin{split}
    W^\phi_{[1.\ell]}=&
    B^\phi_{[0.\ell]}
    -
    \frac{K_\ell B^\delta_{[1.\ell]}}{\lambda^2}
    -\frac{   B^\phi_{[3.\ell]}}{3\lambda^3}.
\end{split}
\end{equation}
This give us for the first order axisymmetric perturbations 
\begin{align}
    \delta_{[1]}(\lambda, y)&=\left(B_{[0.\ell]}^\delta+\frac{B_{[1.\ell]}^\delta}{\lambda}\right)P^2_\ell(y)\\
    W^\phi_{[1]}(\lambda, y)&=\left[  B^\phi_{[0.\ell]}
    -
    \frac{K_\ell B^\delta_{[1.\ell]}}{\lambda^2}
    -\frac{   B^\phi_{[3.\ell]}}{3\lambda^3} \right]\frac{P^1_\ell(y)}{s}
\end{align}
Again  we set the unphysical modes $ B^\phi_{[0.0]}= B^\phi_{[3.0]}=0$, because of  behavior of the angular base functions of $W^\phi_{[1]}$.

Inserting the obtained solutions into  \eqref{ev_lin} yields while using \eqref{rel3b_aLP}
\begin{equation}
    0= \frac{A}{2}\left[\lambda \delta_{[1.\ell],\lambda }\right]_{,\lambda }
    -\left[\lambda^2\delta_{[1.\ell],\lambda }\right]_{,\lambda }
    -\frac{1}{2} \left( \lambda^2W^\phi_{[1.\ell]} \right)_{,\lambda }
\end{equation}
and together with \eqref{sol_d1_ell} and \eqref{sol_w31_ell} this gives us for any $\ell\ge2$ 
\begin{equation}
     0= 
     -B^\phi_{[0.\ell]} \lambda
     +\frac{1}{\lambda^2}\left(\frac{A}{2}B_{[0.\ell]}^\delta-\frac{   B^\phi_{[3.\ell]}}{6}\right)\;\;.
\end{equation}
Hence for any $\ell\ge 2$, 
\begin{equation}\label{constr_B30_Bdelta_1}
    B^\phi_{[0.\ell]} = 0\;\;,\;\;
    B_{[0.\ell]}^\delta = \frac{B^\phi_{[3.\ell]}}{3A}.
\end{equation}
Moreover, inserting the obtained solution into the supplementary equation \eqref{supp_lin} while using \eqref{rel4_aLP}, we find
\begin{equation}\label{eq:supp_S3_lin}
\begin{split}
     \hat S_{3} =& \left\{\frac{B^\phi_{[0,\ell]}}{2}
 -\frac{\ell(\ell+1)B^\delta_{[1.\ell]}}{2\lambda^2}
 +\frac{3AB^\delta_{[1.\ell]} - B^\phi_{[3.\ell]}}{6\lambda^3}
  \right\}\\
  &\times K_\ell \times s(y)P^1_\ell(y)\;.
\end{split}
\end{equation}
Considering \eqref{eq:supp_S3_lin} for the various modes of $\ell$ gives us: $\ell=0$ is trivial because $P^1_0=0$; the $\ell=1$ coefficient vanishes since $K_1=0$. Therefore the coefficients $B^\phi_{[0.1]}$ and $B^\phi_{[3.1]}$ are {\it unconstrained} by the supplementary equation $\hat S_{3}$.  Finally considering $\hat S_{3} = 0$  for the $\ell>1$ coefficients while using \eqref{constr_B30_Bdelta_1} gives us 
\begin{align}
    0=&\ell(\ell+1)B^\delta_{[1.\ell]}
\end{align}
which implies
\begin{equation}
    B^\delta_{[1.\ell]}
    =0 \;\;:\;\;\forall \ell>1\;\;.
\end{equation}

Furthermore, requiring  an asymptotic Bondi frame (a non-rotating inertial observer at large distances), i.e.
\begin{eqnarray}
g_{ab}dx^adx^b\rightarrow -du^2-d\lambda du+\lambda^2q_{AB}dx^Adx^B
\end{eqnarray}
annuls the integration constants,
\begin{equation}
  W^\phi_{[0.1]} =B_{[0.\ell]}^\delta=0.
\end{equation}

From the above requirements, the final solution of the linear perturbations are
\begin{equation}
\begin{split}
    &\delta_{[1]}(y, \lambda) = 0\;\;,\;\;\\
    &W^\phi_{[1]}(y,\lambda ) = -\frac{B}{3\lambda^3}\frac{P^1_\ell(y)}{s} 
    = -\frac{B}{3\lambda^3}\frac{y}{s},
\end{split}
\end{equation}
where we redefined  $B:=B^\phi_{[3,1]}$ for notational  convenience because it is the only remaining integration constant.

\subsection{Quadratic perturbations}\label{sec:sol_quad_pert}

Using the notation of Sec.~\ref{sec:operators}, the relevant  main equations (i.e. only those containing $\gamma_{[2]}$, $R_{[2]}$, $W^y_{[2]}$ and $W_{[2]}$) for the quadratic perturbations are found to be
\begin{subequations}
\begin{align}
    0=&\hat S_{1}(W_{[2]}, W^y_{[2]}) 
    + \frac{B^2 s^2}{2\lambda^6}\left(1-\frac{A}{2\lambda}\right)\label{supp1_quad} \\
    0=&\hat S_{ 2}(W_{[2]}, W^y_{[2]})\label{supp2_quad} \\
   0=& \hat H^{(\gamma)}_{1}(R _{[2]})\label{eq:hyp_E1_2}\\
    0=& \hat H^{(\gamma)}_{2}(R _{[2]}, \gamma_{[2]}, W^y_{[2]})\label{eq:hyp_E2_2}\\
    0=& \hat H^{(\gamma)}_{3}(W_{[2]}, R_{[2]}, \gamma_{[2]}, W^y_{[2]})
    -\frac{ B^2 s^2}{4\lambda^4}\label{eq:hyp_E4_2}\\
    0=& \hat H^{(\gamma)}_{4}(\gamma_{[2]}, W^y_{[2]})
    +\frac{B^2 s^2}{4\lambda^4}\label{eq:hyp_E5_2}
\end{align}
\end{subequations}
The first hypersurface equation \eqref{eq:hyp_E1_2} is readily integrated
%\begin{equation}\label{sol_R1}
%   \xout{ R_{[2]} = C_{R20}(y)+C_{R11}(y)r.}
%\end{equation}
%
\begin{equation}\label{sol_R1}
    R_{[2]} = C_{R20}(y)+C_{R11}(y)\lambda.
\end{equation}
where $C_{R20}(y)$ and $C_{R11}(y)$ are free functions of $y$.
Similarily to \eqref{eq:master_d1}, we can deduce  a master equation for $\gamma_{[2]}$ 
\begin{align}
    0=&\mathcal{M}(\gamma_{[2]}) 
    -s^2  R_{2,\lambda\lambda  y y }
    -\frac{5B^2}{2\lambda^5}s^2.\label{eq:master_g2}
\end{align}
For finding a solution of the remaining fields $\gamma_{[2]}$, $W^y_{[2]}$ and $W_{[2]}$, we need to solve the master equation \eqref{eq:master_g2}. Defining 
\begin{equation}\label{def:H_2}
    \psi_{[2]} = (\lambda\gamma_{[2]})_{,\lambda\lambda }
\end{equation}
with Legendre decomposition
\begin{equation}\label{decomp:H_2}
    \psi_{[2]} = \psi_{[2.\ell]}(\lambda)P^2_ \ell(y)
\end{equation}
while using \eqref{eq:hyp_E1_2} gives us after insertion of \eqref{sol_R1}, \eqref{def:H_2} and \eqref{decomp:H_2}  into \eqref{eq:master_g2}
\begin{equation}\label{master_H2}
\begin{split}
    0 = &
    \left\{-\frac{1}{2}r(A-2\lambda)\frac{d^2\psi_{[2.\ell]}}{d\lambda^2}
    +\left(4r-\frac{A}{2}\right)
    \frac{d\psi_{[2.\ell]}}{dy}\right.\\
    &
    +\left.\left[2-\ell(\ell+1)+\frac{A}{2\lambda}\right]\psi_{[2.\ell]}\right\}P^2_\ell
    %\\
    %&
    -\frac{5B^2}{2\lambda^5}s^2%(1-y^2)
\end{split}
\end{equation}
To fully factor out the Legendre polynomials $P^2_\ell$, we  recall that $P^2_2(y) = 3s^2$. This allows us to write
\begin{equation}\label{master_H2_ell}
\begin{split}
    &0 = 
    \left[\left\{-\frac{1}{2}r(A-2\lambda)\frac{d^2\psi_{[2.\ell]}}{d\lambda^2}
    +\left(4r-\frac{A}{2}\right)
    \frac{d\psi_{[2.\ell]}}{dy}\right.\right.\\
    &
    +\left.\left.\left[2-\ell(\ell+1)+\frac{A}{2\lambda}\right]\psi_{[2.\ell]}\right\}\delta^{\ell^\prime}_
    \ell
    %\\
    %&
    -\frac{5B^2}{6\lambda^5}\delta^{\ell^\prime}_2\right]P^2_{\ell^\prime}(y)
\end{split}
\end{equation}

We can see that  \eqref{master_H2_ell} resembles \eqref{master_H1} if $B=0$. It is in fact a inhomogeneous version of \eqref{master_H1}. We seek solutions of \eqref{master_H2_ell} as a superposition of a homogeneous solution, $\psi^{(hom)}_{[2.\ell]}$ for $B=0$,   and a particular solution $\psi^{(part)}_{[2.\ell]}$ for $B\neq0$, i.e.
\begin{equation}\label{eq:hom_part}
    \psi _{[2.\ell]} = 
    \psi^{(hom)}_{[2.\ell]}+\psi^{(part)}_{[2.\ell]}.
\end{equation}
The homogeneous solution $\psi^{(hom)}_{[2.\ell]}$ will be like \eqref{gen_Sol_psi_LP}.  
Also note that a particular solution  needs to be found for the $\ell=2$ mode, only. We find ${\psi^{(part)}_{[2.2]} = - B^2/(9A\lambda^4)}$. Hence, 
\begin{equation}\label{gen_Sol_psi_LP_quad}
\begin{split}  
    \psi _{[2.\ell]}(\lambda) =& A\left\{\frac{C_{[1.\ell]}P^2_\ell\left(\frac{4\lambda}{A}-1\right)
    +C_{[2.\ell]}Q^2_\ell\left(\frac{4\lambda}{A}-1\right)}{2A-4\lambda}\right\}\\
    &
    +\left(-\frac{ B^2}{9A\lambda^4}\right)\delta^2_\ell.
\end{split}
\end{equation}
It follows by the same regularity arguments like in the discussion for \eqref{gen_Sol_psi_LP} 
that  in order the Weyl curvature scalar $\Psi_0$ does not blow up at the horizon of the unperturbed solution and towards null infinity we must set $C_{[1.\ell]}=C_{[2.\ell]}=0$.

Consequently a solution for the $\psi_{[2.\ell]}$--modes is 
\begin{equation}
     \psi _{[2.\ell]}(\lambda) = \left(-\frac{ B^2}{9A\lambda^4}\right)\delta^2_\ell\;\;.
\end{equation}

Setting 
\begin{equation}\label{def_g2_LP}
    \gamma_{[2]}(\lambda, y) = \gamma_{[2.\ell]}(\lambda) P^2_{\ell}(y),
\end{equation}
we find after integration of \eqref{def:H_2}
\begin{equation}\label{sol_g2_ell}
    \gamma_{[2.\ell]}(\lambda ,y) =
     C^{\gamma}_{[0.\ell]} + \frac{C^{\gamma}_{[1.\ell]}}{\lambda} -\frac{B^2}{54A\lambda^3}\delta_{\ell}^2
\end{equation}
Insertion of \eqref{sol_g2_ell} and \eqref{sol_R1} into \eqref{eq:hyp_E2_2} gives us
\begin{equation}
\begin{split}
    0=&
    \frac{(\lambda^4 W^y_{[2],r})}{2s\lambda^2} 
    +  \frac{sC_{R20,y}}{\lambda^2} 
    %\\
    % &
    +\left(\frac{d\gamma_{[2.\ell]}}{d\lambda}\right)\frac{1}{s}\frac{d}{dy}\left[s^2P^2_\ell\right].
\end{split}
\end{equation}
using \eqref{rel1_aLP}  we find 
 \begin{equation}
\begin{split}
    0=&
    \left(\lambda^4 \frac{W^y_{[2],r}}{s}\right)_{,\lambda}+2sC_{R20,y}
   %\\&    
    -2\lambda^2K_\ell\left(\frac{d\gamma_{[2.\ell]}}{d\lambda}\right) P^1_\ell(y)
\end{split}
\end{equation}
which indicates that the angular behaviour of $W_{[2]}^y/s$  and $sC_{R20,y}$ are
 dictated by the associated Legendre polynomials $P^1_\ell(y)$.
 As of \eqref{rel3a_aLP}, we set (note $P_\ell(y)=P^0_\ell(y)$)
 \begin{align}
     R_{[2]}(\lambda , y) =&R_{[2.\ell]}(\lambda)P_\ell(y) 
     =\Big[C^R_{[20.\ell]}+ C^R_{[21.\ell]}\lambda\Big]P_\ell(y),
     \label{def_R2_LP}\\
     W^y_{[2]}(\lambda , y) =& W^y_{[2.\ell]}(\lambda)s(y) P^1_\ell(y)\label{def_Wy2_LP}
\end{align}
This gives us 
\begin{equation}\label{W_2ell_ode}
\begin{split}
    0=&
    \frac{d}{d\lambda}\left(\lambda^4 \frac{d}{d\lambda}W^y_{[2.\ell]}\right)
    -2C^R_{[20.\ell]}
   %\\&    
    -2\lambda^2K_\ell \left(\frac{d\gamma_{[2.\ell]}}{d\lambda}\right)\;\;,
\end{split}
\end{equation}
Integrating \eqref{W_2ell_ode} yields
\begin{equation}\label{sol_Wy2_ell}
    \begin{split}
        W^y_{[2.\ell]}=& C^y_{[0.\ell]}
        +\frac{K_{\ell}C^\gamma_{[1.\ell]}
        -C^R_{[20.\ell]}}{\lambda^2}
        -\frac{C^y_{[3.\ell]}}{3\lambda^3}
        %+\frac{K_{\ell}B^2}{36A\lambda^4}\delta^{2}_{\ell} 
        -\frac{B^2}{9A\lambda^4}\delta^{2}_{\ell}
        \end{split}
\end{equation}
where we set the integration constants ${C^y_{[0.0]}=C^y_{[3.0]}=0}$, because $P^1_0(y)=0$.
Considering \eqref{eq:hyp_E5_1} with \eqref{def_g2_LP}, \eqref{def_Wy2_LP}, \eqref{rel3b_aLP} 
and $s^2 = P^2_\ell(y)/3$
gives us
\begin{equation}
 \left[\lambda^2\left(1-\frac{A}{2\lambda}\right)\gamma_{[2.\ell],r}\right]_{,\lambda}
    =
    \frac{1}{2} \left(  \lambda^2W^y_{[2.\ell]} \right)_{,\lambda}
    +\frac{B^2 }{12\lambda^4} \delta^2_\ell
\end{equation}
so that after insertion of \eqref{sol_g2_ell} and \eqref{sol_Wy2_ell}, we obtain
\begin{equation}
    \begin{split}
        %0&=
        %-
        \lambda C^y_{[0.\ell]}&=
        \frac{A}{2\lambda^2}\left(C^\gamma_{[1.\ell]}-\frac{C^y_{[3.\ell]}}{3A}\right)
        %\\
        %&
        +\frac{B^2}{9A\lambda^3}\left(1+\frac{K_\ell}{4}\right)\delta^2_\ell
    \end{split}
\end{equation}
implying  for any $\ell\ge2$
\begin{equation}\label{constr_C30_Bdelta_1}
    C^y_{[0.\ell]} = 0\;\;,\;\;
   C^y 
    _{[3.\ell]} = 3A C_{[1.\ell]}^\gamma %= \frac{}{3A}
\end{equation}
Next, proceed with the hypersurface equation \eqref{eq:hyp_E4_2} for $W_{[2]}$. Insertion of  \eqref{def_g2_LP}, \eqref{def_R2_LP} and  \eqref{def_Wy2_LP}   into \eqref{eq:hyp_E4_2} gives us
\begin{equation}
    \begin{split}
       &(\lambda W_{[2]})_{,\lambda}=\left\{
       -\left[\left(1-\frac{A}{2\lambda}\right)(\lambda R_{[2.\ell]})_{,\lambda}\right]_{,\lambda}
       \right.\\
       &+\ell(\ell+1)\left.\left[
       \frac{R_{[2.\ell]}}{\lambda}
       +\frac{(\lambda^4 W^y_{[2.\ell]})_{,\lambda}}{2\lambda^2}
       -K_\ell\gamma_{[2,\ell]}\right]   
       \right\}P_\ell^0(y)\\
       &\qquad
       -\frac{ B^2 s^2}{4\lambda^4}
    \end{split}
\end{equation}
 Using
 \begin{eqnarray}\label{sin2_LP_l0}
   s^2 = 1-y^2 = \frac{2}{3}[P^0_\ell(y) - P^0_2(y)]
 \end{eqnarray}
 as well as setting
 \begin{equation}
     W_{[2]}(\lambda ,y ) = W_{[2.\ell]}(\lambda)P_\ell^0(y)
 \end{equation}
 yields
 \begin{equation}
    \begin{split}
       &(\lambda W_{[2.\ell]})_{,\lambda}= 
       -\left[\left(1-\frac{A}{2\lambda}\right)(\lambda R_{[2.\ell]})_{,\lambda}\right]_{,\lambda}
       \\
       &+\ell(\ell+1) \left[
       \frac{R_{[2.\ell]}}{\lambda}
       +\frac{(\lambda^4 W^y_{[2.\ell]})_{,\lambda}}{2\lambda^2}
       -K_\ell\gamma_{[2,\ell]}\right]       
       \\
       &\qquad
       -\frac{ B^2 }{6\lambda^4}
       (\delta^0_\ell- \delta^2_\ell)
    \end{split}
\end{equation}
Since $R_{[2.\ell]}$, $W^y_{[2.\ell]}$ and $\gamma_{[2.\ell]}$ are known, we find after integration
\begin{equation}\label{eq:W2_ell_sol}
    \begin{split}
        W_{[2.\ell]}&=
        -K_\ell C^R_{[21.\ell]}
        -\ell(\ell+1)K_\ell C^\gamma_{[0.\ell]}
        +\frac{C^W_{[1.\ell]}}{\lambda}
        +\frac{AC^R_{[20.\ell]}}{2\lambda^2}
        \\
        &
       +\frac{\ell(\ell+1)C^y_{[3.\ell]}}{6\lambda^2}  
       +\left(\frac{2 B^2}{9 A\lambda^3}
          -\frac{ B^2 }{18\lambda^4}
          \right)\delta^2_\ell
        +\frac{ B^2 }{18\lambda^4}
       \delta^0_\ell%- \delta^2_\ell)
    \end{split}
\end{equation}
where $C^W_{[1.\ell]}$ are integration constants.

Calculation of \eqref{supp1_quad} and \eqref{supp2_quad} while using \eqref{def_R2_LP}, \eqref{def_Wy2_LP}, \eqref{sin2_LP_l0}, \eqref{aLP_ODE}( for $\hat{m}=0$) and \eqref{rel5_aLP} gives us 
\begin{align}
    0=&
          \left(1-\frac{A}{2\lambda}\right)
          %\left[
        \left(\lambda^2W_{[2.\ell],r}  
        + \frac{AR_{[2.\ell]}}{\lambda}\right)_{,\lambda}
       % +\frac{2B^2}{3\lambda^4}(\delta^0_\ell-\delta^2_\ell)\right ]\lambda
    \nonumber\\
    &-  \ell(\ell+1)\left(  W_{[2.\ell] }
    +\frac{A}{2}W^y_{[2,\ell]}\right)\\
       0&=
        \frac{1}{2\lambda^2}  \left(1-\frac{A}{2\lambda}\right) (\lambda^4 W^y_{[2.\ell]})_{,\lambda}  
     -\frac{1}{2}W_{[2.\ell],r}
     +W^y_{[2.\ell]}
\end{align}
and insertion of the respective coefficient solutions \eqref{def_R2_LP},\eqref{sol_Wy2_ell} and \eqref{eq:W2_ell_sol} yields 
%\begin{widetext}
\begin{align}
0&=
\ell(
 \ell+1)\left\{
 -\frac{C^W_{[1.\ell]}}{\lambda}
 +K_\ell\left[
  C^\gamma_{[1.\ell]}-C^R_{[21.\ell]}
 \right]
 \right.\nonumber
 \\
 &\left.\qquad
 +\frac{C^y_{[3.\ell]} - 3AC^\gamma_{[1.\ell
 ]}}{6\lambda^2}
 \right\}
\\
0&=
\frac{C^W_{[1.\ell]}}{2\lambda^2}
+\frac{(C^y_{[3.\ell]}-3AC^\gamma_{[1.\ell]} )K_\ell}{6\lambda^3}\;\;,\qquad
\forall\ell\ge 1
\end{align}

Therefore,
\begin{subequations}\label{cond_suppl_quad_pert}
  \begin{align}
C^W_{[1.\ell]}=&0\;\;,\;\;\forall\ell\ge 1\\
C^y_{[3.\ell]}=&3AC^\gamma_{[1.\ell]}\;\;,\;\;\forall\ell\ge 2\label{cond_Cy_3L_supp}\\
    C^R_{[21.\ell]}=&C^\gamma_{[1.\ell]}
    \;\;,\;\;\forall\ell\ge 2
\end{align}  
\end{subequations}
Note, \eqref{cond_Cy_3L_supp} is consistent with \eqref{constr_C30_Bdelta_1}. 
The requirement of an asymptotic inertial observer leads to  
\begin{align}
    %C^\gamma_{[1.\ell]}=
    C^{\gamma}_{[0.\ell]} =
    C^R_{[21.\ell]} =0
\end{align}
 which gives with \eqref{cond_suppl_quad_pert}  that
$
    C^\gamma_{[1.\ell]}=
    C^y_{[3.\ell]} =0.
$ Thus, redefining  $C:=C^W_{[1.1]}$, the quadratic perturbations are \begin{eqnarray}
     \gamma_{[2]}(\lambda ,y)&=&
      \left(
     -\frac{B^2}{54A\lambda^3}\delta^2_\ell\right)P^2_\ell(y)\\
     R_{[2]}(\lambda ,y)&=& 0\\    
     W^y_{2}(\lambda ,y)&=&
     \left( 
        -\frac{B^2}{9A\lambda^4}\delta^{2}_{\ell}  \right)s(y)P^1_\ell(y)\\
     W_{[2]}(\lambda ,y)&=&\frac{C}{\lambda}
     +\frac{ B^2 }{18\lambda^4} 
       +\left(\frac{2 B^2}{9 A\lambda^3}
          -\frac{ B^2 }{18\lambda^4}
          \right)P^0_2(y)\;.
          \nonumber
          \\
 \end{eqnarray}

\subsection{Third order perturbations}

Similarly, expressions for the higher order perturbations quantities $f_{[i]}$ can be obtained using the same \blue{procedure} as in the previous sections. In this and in the next subsection we show the fundamental results without repeating  intermediate steps.

The relevant equations for the third perturbations are 
\begin{align}
    0=&\hat S_{3}(\delta_{[3]}, W^\phi_{[3]})
    -\frac{B^3 s^4}{6A\lambda^6}
    \label{supp_cub} \\
    0=&\hat H^{(\delta)}_{1}(\delta_{[3]}, W^\phi_{[3]})
    -\frac{B^3 s^4}{6A\lambda^6}
    \label{hyp_Wphi_cub}\\    
    0=&\hat H^{(\delta)}_{2}(\delta_{[3]}, W^\phi_{[3]})
        +\frac{2B^3y s^2}{3A\lambda^5}
\label{ev2_cub}
\end{align}

Similarily to  \eqref{eq:master_d1}, we can deduce a master equation for $\delta_{[3]}$  
\begin{subequations}
\begin{align}
    0=&\mathcal{ M}(\delta_{[3]})
    -\frac{40B^3}{3A\lambda^6} s^2 y
    %+\sin^2\theta\cos\theta
    \label{eq:master_d3}
\end{align}
\end{subequations}
Using $P^2_3(y) = 15ys^2$ and following the steps of Sec.~\ref{sec:sol_lin_pert}, we find
\begin{align}
    \delta_{[3]}(\lambda,y) = &\left(-\frac{B^3}{ 162A^2 \lambda^4}\right) P^2_3(y)\\
    W^\phi_{[3]}(\lambda,y)=& \left[-\frac{D }{3\lambda^3} -\frac{2B^3}{135A\lambda^6} \right]\frac{P^1_1(y)}{s(y)}\nonumber\\
    &+\left[\frac{B^3}{ 405A\lambda^6} - \frac{4B^3}{81A^2\lambda^5} \right]\frac{P^1_3(y)}{s(y)}
\end{align}
where $D$ is the only free new remaining integration constant  that appears at this order.

\subsection{Fourth order perturbations}\label{sec:sol_quartic}
Here the relevant  main equations are those containing $\gamma_{[4]}$, $R_{[4]}$, $W^y_{[4]}$ and $W_{[4]}$ 
which are
\begin{subequations}
\begin{align}
    0=&\hat S_{1}(W_{[4]}, W^y_{4]}) 
    + \left(\frac{14}{9A\lambda}-\frac{1}{12\lambda^2}-\frac{35}{6A^2}  \right)\frac{B^4s^4}{3\lambda^8}
    \nonumber\\
    &+\left[\left(1-\frac{A}{2\lambda}
      \right)D -\frac{CB}{2A}
      +\left(\frac{16}{A\lambda^2}-\frac{7}{3\lambda^3}\right)\frac{B^3}{9A}\right]\frac{Bs^2}{\lambda^6}
      \nonumber\\
      &
      -\frac{8B^4}{27\lambda^8A^2}
      +\frac{CB^2}{3A\lambda^6}
    \label{supp1_quart} \\
    0=&\hat S_{2}(W_{[4]}, W^y_{[4]})
    +\left[\frac{(7A+120\lambda)s^2}{12\lambda}-8\right]\frac{B^4ys}{9A^2 \lambda^7}\nonumber\\
    &
    +\frac{2ysCB^2}{3A\lambda^5}\\
    %\label{supp2_quart} 
%\end{align}
%\begin{align}
0=& 
\hat H^{(\gamma)}_{1}(R_{[4]})
    - \frac{B^4s^4}{18A^2\lambda^8}
   \label{eq:hyp_E1_4}
   \\
    0=& \hat H^{(\gamma)}_{2}(R_{[2]}, \gamma_{[2]}, W^y_{[2]})
    +\frac{B^4 ys^3}{27 A^2 \lambda^7}
    \label{eq:hyp_E2_4}\\
    0=&
    \hat H^{(\gamma)}_{3}(W_{[2]}, R_{[2]}, \gamma_{[2]}, W^y_{[2]})
    +\frac{(A-14r)B^4s^4 }{36A^2 \lambda^7}
    \nonumber\\
    &+\frac{2B^4s^2 }{9A^2 \lambda^6} +\frac{DBs^4}{2\lambda^4} 
    \label{eq:hyp_E4_4}\\
    0=& 
    \hat H^{(\gamma)}_{4}(\gamma_{[2]}, W^y_{[2]})
    +\left(14+\frac{A}{2r}\right)\frac{B^4 s^4}{9A^2\lambda^6}
    \nonumber\\
    &
    +\left(\frac{B^2C}{2A\lambda^4}
    -\frac{BD}{2\lambda^4}
    -\frac{38B^2}{27A^2\lambda^6}\right)s^2
    \label{eq:hyp_E5_4}
\end{align}
\end{subequations}
The first hypersurface equation \eqref{eq:hyp_E1_4} is readily integrated
\begin{equation}\label{sol_R4}\begin{split}
    R_{[4]}(\lambda, y) = &E_{R0}(y)+E_{R1}(y)\lambda  -\frac{s^4B^4}{1080A^2\lambda^5} 
\end{split}
\end{equation}
or expressing in terms of the Legendre polynomials $P^0_\ell(y)$
\begin{equation}\label{sol_R4_LP}
\begin{split}
    R_{[4]}(\lambda,  y) = &\left(E^R_{[0.\ell]}+E^R_{[1.\ell]}\lambda\right)P^0_\ell(y)\\
    &   -\frac{B^4}{135A^2\lambda^5}\left(\frac{P^0_0(y)}{15}
    -\frac{2P^0_2(y)}{21}
    +\frac{P^0_4(y)}{35}\right)
\end{split}
\end{equation}

Similarily to  \eqref{eq:master_g1} we can deduce a master equation  for  $\gamma_{[4]}$ 
\begin{subequations}
\begin{align}
    0=&\mathcal{M}(\gamma_{[4]}) 
    -  s^2R_{[4],\lambda\lambda y y }
        - \frac{5 B^2 C s^2}{A\lambda^5}
        - \frac{5 B D s^2}{\lambda^5}
        \nonumber\\
&+ \left[358-s^2\left(397+\frac{A}{\lambda} \right)\right]\frac{B^4s^2}{9A^2\lambda^7}
    \label{eq:master_g4}
\end{align}
\end{subequations}
Using the methods of Sec.~\ref{sec:sol_quad_pert} together with the inverted Legendre relations
\begin{subequations}
    \begin{align}
        1=& P^0_0(y) = -\frac{P^1_1(y)}{s}\\
        y=& P^0_1(y) = -\frac{P^1_2(y)}{3s}\\
        y^2 =&\frac{1}{3}-\frac{2}{3}P^0_2(y) 
        =  1-\frac{1}{3}P^2_{2}(y)
        \\
        y^3=&-\frac{2P^1_4(y)}{35s} + \frac{P^1_2(y)}{7s} \\
        y^4 =& 
        \frac{1}{5}-\frac{4P^0_2(y)}{7}+\frac{8P^0_4(y)}{35}
        =1- \frac{8}{21}P^2_2(y)-\frac{2}{105}P^2_4(y)
    \end{align}
\end{subequations}
we deduce the following solution for the fourth order  perturbation 
\begin{subequations}
\begin{align}
   R_{[4]} =& -\frac{B^4}{135A^2\lambda^5}\left(\frac{P^0_0}{15}
    -\frac{2P^0_2}{21}
    +\frac{P^0_4}{35}\right)\\
\gamma_{[4]} =&  \left(\frac{BD}{27A\lambda^3} 
- \frac{B^2C}{ 27 A^2 \lambda^3}
-\frac{B^4}{1134A^2\lambda^6}\right)P^2_2\nonumber\\
     &
     +\left(\frac{B^4}{405A^3\lambda^5 } 
     + \frac{B^4}{17010A^2\lambda^6}\right)P^2_4\\
     W^y_{[4]}=&
     \left(\frac{2BD}{ 9A\lambda^4}
     - \frac{2B^2C}{9A^2\lambda^4}
     - \frac{2B^4}{ 2835A^2\lambda^7}\right)P^1_2
     \nonumber\\
     &
     +\left(\frac{2B^4}{81A^3\lambda^6}
     + \frac{B^4}{ 4725 A^2\lambda^7}\right)P^1_4\\
   W_{[4]}=&\frac{E }{\lambda} 
   -\frac{BD}{9\lambda^4}
   +\frac{4B^4}{ 405A^2\lambda^6} 
   - \frac{B^4}{675A\lambda^7}
   \nonumber\\
   +\bigg( &\frac{2B^4}{ 945A\lambda^7} 
-\frac{2B^4}{81A^2\lambda^6} 
   + \frac{4B^2C}{9A^2\lambda^3} 
   -\frac{4BD}{9A\lambda^3} 
   + \frac{BD}{ 9\lambda^4}\bigg)P^0_2\nonumber\\
   &+\left(-\frac{ 8B^4}{81A^3\lambda^5}
   + \frac{2B^4}{ 135A^2\lambda^6} 
   - \frac{B^4}{ 1575A\lambda^7}\right)P^0_4
\end{align}
\end{subequations}
Note that $E$ is the only remaining new integration constant, all other vanish because of the reasons mentioned in Sec.~\ref{sec:sol_quad_pert}.

\subsection{Perturbations in terms of Komar quantities}
\label{sec:combine_Komar}
The solution of the perturbation involve the free integration constants $A, B, C, D$ and $E$. These free constants determine the Komar mass, $K_m$, and the Komar angular momentum, $K_L$, which can be found by calculation of \eqref{eq:komm} and \eqref{eq:Lang}
\begin{align}
m:=K_m &= \frac{A}{4} - \frac{C}{2}\varepsilon^2
-\frac{E}{2}\varepsilon^4+O(\varepsilon^5)
\label{Komar_m_4th}\\
L:=K_L&
=-\frac{B}{6}\varepsilon +\frac{D}{6}\varepsilon^3+ O(\varepsilon^5)
\label{Komar_L_4th}
\end{align}
If $\varepsilon=0$, $K_m=A/4$ corresponds to the mass $m_0$ of the unperturbed system. 
Furthermore, we can see that $L=O(\varepsilon)$. 
This allows us relate $\varepsilon$ with the angular momentum $L$ of the system. To do that we have to solve the cubic equation 
\begin{equation}\label{cubic_eps_L}
   0=\frac{D}{6}\varepsilon^3-\frac{B}{6}\varepsilon  +L
\end{equation}
for $\varepsilon$. This equation also shows that in order to make the substitution of  $\varepsilon$ by $L$, we seek the solution  $\varepsilon(L) = O(L)$. The root of \eqref{cubic_eps_L} which fulfils this requirement is 
\begin{equation}\label{subs_eps}
    \varepsilon = -\frac{6}{B}L-\frac{216D}{B^4}L^3 + O(L^5)
\end{equation}
Subsequent insertion of this expansion into \eqref{Komar_m_4th} and solving for $A$ gives us
\begin{equation}\label{A_subs_m}
    A = 4m + \frac{72C}{B^2}L^2 + 2592\frac{EB-2CD}{B^5}L^4 + O(L^6)
\end{equation}

The relations \eqref{subs_eps} and \eqref{A_subs_m} allow us to substitute $A$ and $\varepsilon$, by the physical quantities $m$ and $L$.
Insertion of \eqref{subs_eps} and \eqref{A_subs_m} into the solution of the perturbations and subsequent expansion up to $O(L^4)$ allows us to eliminate the integration constants $C,D$ and $E$ from the perturbations. Meaning all integration constants are absorbed into the Komar mass, $m$ and the Komar angular momentum, $L$. Thus the final solution is uniquely described by the two physical quantities $m$ and $L$. This gives us 
\begin{widetext}
\begin{subequations}\label{eq:Pert_sol_m_L}
    \begin{eqnarray}
        R(\lambda, y)&=&
        \lambda
        -\left(\frac{P^0_0}{5}-\frac{2P^0_2}{7}+\frac{3P^0_4}{35}\right)\frac{L^4}{5m^2\lambda^5}
        +O(L^6)\\
        W(\lambda, y) &=& 1-\frac{2m}{\lambda}
        +\left[\frac{2}{\lambda^4}
        +\left(\frac{2}{m\lambda^3}
        -\frac{2}{\lambda^4}\right)P^0_2\right]L^2
        +\left[\frac{4}{5m^2\lambda^6}
           -\frac{12}{25\lambda^7}
        +\left(\frac{24}{35m\lambda^7}-\frac{2L^4}{m^2\lambda^6}\right)P^0_2\right.\nonumber\\
        &&\qquad
        +\left.
        \left(-\frac{2}{m^3\lambda^5}
              +\frac{6}{5m^2\lambda^6}
              -\frac{36}{175m\lambda^7}\right)P^0_4
        \right]
        L^4 + O(L^6)\\
        W^y(\lambda, y)
        &=&
        \left(-\frac{1}{m\lambda^4}(sP^1_2)\right)L^2
        +\left[-\frac{2  }{35m^2\lambda^7}(sP^1_2)
              +\left(\frac{1}{2m^3\lambda^6}
               +\frac{3 }{175m^2\lambda^7}\right)(sP^1_4)\right]L^4
               + O(L^6)\\
        W^\phi(\lambda, y)
        &=&
        \left(-\frac{2}{\lambda^3}\frac{P^1_1}{s}\right)L
        +\left[\frac{4}{5m \lambda^6}\frac{P^1_1}{s}
          +\left(\frac{2}{3m^2\lambda^5}-\frac{2}{15 m\lambda^6}\right)\frac{P^1_3}{s}\right]L^3
        + O(L^5)\\        
         \gamma(\lambda, y)&=&
         \left(-\frac{1}{6m\lambda^3}P^2_2\right)L^2
         +\left[-\frac{1}{14m^2\lambda^6}P^2_2
         +\left(\frac{1}{20m^3\lambda^5}
         +\frac{1}{210m^2\lambda^6}\right)P^2_4\right]L^4
         +O(L^6)\\
         \delta(\lambda, y)&=&
         \left(\frac{1}{12m^2\lambda^4}P^2_3\right)L^3
         +O(L^5)
    \end{eqnarray}
\end{subequations}
We see in \eqref{eq:Pert_sol_m_L} that the perturbations are determined by the mass and angular momentum, i.e. the solution has two hairs. To show that this solutions represents the Kerr solution in affine-null coordinates, we introduce the specific angular momentum, $a:=L/m$. In terms of $a$, \eqref{eq:Pert_sol_m_L} read after changing to the angular coordinate $\theta$
\begin{subequations}\label{eq:Pert_sol_m_a}
    \begin{eqnarray}
        R(\lambda, \theta)&=&
        \lambda
        - \frac{3 m^2\sin^4\theta}{40\lambda^5}a^4
        +O(a^6)\\
        W(\lambda, \theta) &=& 1-\frac{2m}{\lambda}
        +\left[\frac{2m}{\lambda^3}
        +\left(-\frac{3m}{\lambda^3}
        + \frac{3m^2}{\lambda^4}\right)\sin^2\theta \right]a^2
        \nonumber\\
        &&
        +\left[
        -\frac{2m}{\lambda^5}
        +\left(\frac{10m}{\lambda^5} - \frac{3m^2}{\lambda^6}\right)\sin^2\theta
        +\left(-\frac{35m}{4\lambda^5} 
        + \frac{ 21m^2}{ 4\lambda^6} 
        - \frac{ 9m^3}{10\lambda^7} 
        \right)\sin^4\theta
        \right]
        a^4 + O(a^6)\\
        W^\theta(\lambda, \theta)
        &=&\left\{
         -\frac{3m}{\lambda^4}a^2        
        +\left[ \frac{5m  }{\lambda^6}  
              -\left(\frac{35m}{4 \lambda^6}
               +\frac{3 m^2 }{10\lambda^7}\right)\sin^2\theta\right]a^4
        \right\}\sin\theta\cos\theta
               + O(a^6)\\
        W^\phi(\lambda, \theta)
        &=&
         \frac{2m}{\lambda^3}a
        +\left[
          -\frac{4m}{\lambda^5}
          +\left(\frac{5m}{\lambda^5}-\frac{m^2}{ \lambda^6}\right)\sin^2\theta\right]a^3
        + O(a^5)\\        
         \gamma(\lambda, \theta)&=&
         \left(-\frac{m\sin^2\theta}{2\lambda^3} \right)a^2
         +\left[\frac{9m\sin^2\theta}{4\lambda^5}
         +\left(-\frac{21m}{ 8\lambda^5} - \frac{m^2}{ 4\lambda^6}\right)\sin^4\theta\right]a^4
         +O(a^6)\\
         \delta(\lambda, \theta)&=&
         -\frac{5m \cos\theta\sin^2\theta}{4\lambda^4}a^3
         +O(a^5)
    \end{eqnarray}
\end{subequations}
%\end{widetext}
Comparing with \cite{2007PhRvD..75d4003B}, we find agreement for $R$ which corresponds to their areal coordinate $r$.
Calculation of the metric components $g_{ab}$ using \eqref{eq:Pert_sol_m_a} 
gives us
%\begin{widetext}
\begin{subequations}\label{eq:srot_metric_final}
\begin{eqnarray}
g_{uu}(\lambda, \theta)&=&-1+\frac{2m}{\lambda}
+\left[
  \left(\frac{3m}{\lambda^3} + \frac{m^2}{\lambda^4}\right)
  \sin^2\theta
   - \frac{2m}{\lambda^3}
   \right]a^2
   \nonumber\\
   &&
+\left[
\frac{2m}{\lambda^5}
-\left(\frac{10m}{\lambda^5} + \frac{ 4m^2}{ \lambda^6}
\right)\sin^2\theta 
+
\left(\frac{35m}{4\lambda^5} 
+ \frac{23m^2}{ 4\lambda^6}
+ \frac{9m^3}{10\lambda^7}\right)\sin^4\theta 
\right] a^4 + O(a^6)\\
g_{u\lambda}(\lambda, \theta)&=&-1\\
g_{u\theta}(\lambda, \theta)&=&
  \left\{
  \left(\frac{3m}{\lambda^2}  \right)a^2
  +\left[-\frac{5m}{\lambda^4} 
  +\left(
  \frac{35m}{\lambda^4}
  +\frac{23m^2}{10\lambda^5}
  \right)\sin^2\theta\right]a^4 \right\}\sin\theta\cos\theta
  + O(a^6)\\
g_{u\phi}(\lambda, \theta)&=&\left\{
      \left(-\frac{2m}{\lambda}\right)a
     +\left[\frac{4m}{\lambda^3} 
  -\left(
   \frac{5m}{\lambda^3}
  +\frac{m^2}{\lambda^4}
  \right)\sin^2\theta\right]a^3 \right\}\sin^2\theta  
     +O(a^5)\\
g_{\theta\theta}(\lambda, \theta)&=&\lambda^2  
+ \left(-\frac{m\sin^2\theta}{\lambda}\right)a^2 
+\left[\frac{9m}{2\lambda^3}\sin^2\theta
  -\left(
   \frac{21m}{4\lambda^3}
  +\frac{3m^2}{20\lambda^4}
  \right)\sin^4\theta\right]a^4 
+ O(a^6)\\
g_{\theta\phi}(\lambda, \theta)&=&
\left(-\frac{5m\sin^3\theta\cos\theta}{2\lambda^2}\right)a^3
+O(a^5)\\
g_{\phi\phi}(\lambda, \theta)&=&
\left\{\lambda^2  
+ \left(\frac{m\sin^2\theta}{\lambda}\right)a^2 
+\left[-\frac{9m}{2\lambda^3}\sin^2\theta
  +\left(
   \frac{21m}{4\lambda^3}
  +\frac{17m^2}{20\lambda^4} \right)
  \sin^4\theta\right]a^4 
     \right\}\sin^2\theta
+ O(a^6)
\end{eqnarray}
\end{subequations}
\end{widetext}
Eqs.\eqref{eq:srot_metric_final} constitute our final expression for the slowly rotating stationary and axially symmetric (Kerr) metric adapted to null coordinates which asymptotically match an inertial Bondi frame. At difference of all previous approaches, it was obtained as {\it an explicit solution} of the Einstein equations.
After comparison with \cite{2007PhRvD..75d4003B}, we find agreement up to a typo in their equation for $g_{\theta\phi}$. We also note  care should be taken when comparing \cite{2007PhRvD..75d4003B}'s expressions with ours. First, \cite{2007PhRvD..75d4003B} present a Bondi-Sachs form of the metric, while we have an affine-null metric approaching a Bondi frame, the difference is in the choice of radial coordinate, and the two agree only up to $O(\lambda^{-4})$ with one another. Second, \cite{2007PhRvD..75d4003B} make a large $\lambda$ expansion while we make a small $a$ expansion, this results in  powers of $\lambda^{-k}$ absorbed by order symbols in 
\cite{2007PhRvD..75d4003B}.
A slowly rotating version of the Kerr metric in null affine coordinates at second order in $a$ was also obtained by Dozmorov 
who made a null tetrad rotations starting with the Kerr metric as expressed in Boyer-Lindquist coordinates \cite{1975Fiz....18...95D}. In the next section we show an alternative procedure to recover the slowly rotating Kerr metric components as expressed in \eqref{eq:srot_metric_final} by doing appropriate coordinates transformations.

\section{Approximated affine-null metric derived from the Kerr-metric}\label{sec:affine_null_kerr_trafo}
Here, starting with the Kerr metric expressed in Boyer-Lindquist coordinates (BL) $\{\hat t, \hat r,\hat \theta,\hat\phi\}$, we present an explicit transformation to affine-null coordinates up to fourth  order in a.
The Kerr metric in BL coordinates reads:
\begin{equation}
    ds^2=g_{\hat{t}\hat{t}}d\hat{t}^2+g_{\hat{t}\hat{\phi}}d\hat{t}d\hat{\phi}+g_{\hat{r}\hat{r}}d\hat{r}^2+g_{\hat{r}\hat{r}}d\hat{r}^2+g_{\hat{\theta}\hat{\theta}}d\hat{\theta}^2+g_{\hat{\phi}\hat{\phi}}d\hat{\phi}^2;
\end{equation}
with
\begin{eqnarray}
 g_{\hat{t}\hat{t}}&=&-\left(1-\frac{2m\hat{r}}{\Sigma}\right),\\
 g_{\hat{t}\hat{\phi}}&=&-\frac{2ma\hat{r}\sin^2\hat{\theta}}{\Sigma},\\
 g_{\hat{r}\hat{r}}&=&\frac{\Sigma}{\Delta},\\
g_{\hat{\theta}\hat{\theta}}&=&\Sigma,\\
 g_{\hat{\phi}\hat{\phi}}&=&\left(\hat{r}^2+a^2+\frac{2ma^2\hat{r}\sin^2\hat{\theta}}{\Sigma}\right)\sin^2\hat{\theta},
\end{eqnarray}
with $\Delta=\hat{r}^2-2m\hat{r}+a^2$ and $\Sigma=\hat{r}^2+a^2\cos^2\hat{\theta}$.
The $u$ null coordinate must satisfy the eikonal equation,
\begin{equation}
g^{ab}\nabla_a u\nabla_b u=0,\label{eq:eqikgen}
\end{equation}
Inspired by \cite{2007PhRvD..75d4003B}, we propose the following expansion for $u$,
\begin{equation}
    u=\hat{t}-\hat{r}-2m\ln\left(\frac{\hat{r}-2m}{2m}\right)+\sum^\infty_{i=1} f_{i}(\hat{r},\hat{\theta})a^i.\label{eq:uasymp}
\end{equation}
Note that for $a=0$ this expression reduces to the standard outgoing Schwarzschild null coordinate.
By replacing \eqref{eq:uasymp} into \eqref{eq:eqikgen}, we obtain a set of differential equations for $f_i(\hat{r},\hat{\theta})$ that can be solved iteratively. Conserving terms up to fourth order in $a$ we find that  only the even coefficients $f_{2n}(\hat{r},\hat{\theta})$ are non--vanishing with:

\begin{eqnarray}   f_2(\hat{r},\hat{\theta})&=&\frac{5\hat{r}-2m}{4\hat{r}(2m-\hat{r})}+\frac{\cos2\hat{\theta}}{4\hat{r}}-\frac{\ln(1-\frac{2m}{\hat{r}})}{2m},\\
f_4(\hat{r},\hat{\theta})&=&\frac{(2\hat{r}+m)}{16\hat{r}^4}\sin^4(2\hat{\theta})-\frac{3\ln(1-\frac{2m}{\hat{r}})}{8m^3}\nonumber\\
&&-\frac{4m^2-9m\hat{r}+3\hat{r}^2}{4m^2\hat{r}(\hat{r}-2m)^2},
\end{eqnarray}

Similarly, affine-null coordinates $\{\lambda,\theta,\phi\}$ can be obtained from the requirements 
\begin{subequations}\label{eq:conditions}
\begin{eqnarray}
g^{ab}\nabla_a u\nabla_b\lambda&=&-1,\\ g^{ab}\nabla_a u\nabla_b\theta&=&g^{ab}\nabla_au\nabla_b\phi=0,
\end{eqnarray}
\end{subequations}
by assuming relations of the form:
\begin{eqnarray} \lambda&=&\hat{r}+\sum^\infty_{i=1}\hat\Lambda_i(\hat{\theta},\hat{r})a^i,\\
\theta&=&\hat{\theta}+\sum^\infty_{i=1}\hat\Theta_i(\hat{\theta},\hat{r})a^i,\\
\phi&=&\hat{\phi}+\sum^\infty_{i=1}\hat\Phi_i(\hat{\theta},\hat{r})a^i,
\end{eqnarray}
and replacing into the set \eqref{eq:conditions}, the coefficients functions $\hat\Lambda_i,\hat\Theta_i,\hat\Phi_i$ can be obtained in the same way as $u$. After that, the resulting relations can be inverted in order to express the BL coordinates in terms of the affine-null  coordinates. Following these steps up to fourth order, the final transformation coordinates reads:
\begin{widetext}
\begin{eqnarray}
    \hat{t}&=&u+\lambda+2m\ln(\frac{\lambda}{2m}-1)+\left[\frac{\ln(1-\frac{2m}{\lambda})}{2m}+{\frac {3\,m\cos \left( 2\,\theta \right) +4\,\lambda-3\,m}{ \left( 4
\,\lambda-8\,m \right) \lambda}}
\right]a^2\nonumber\\
&&+\left[-{\frac {m \left( 175\,{\lambda}^{2}-224\,m\lambda-72\,{m}^{2}
 \right)  \left( \cos \left( 2\,\theta \right)  \right) ^{2}}{320\,
 \left( \lambda-2\,m \right) ^{2}{\lambda}^{4}}}-{\frac {m \left( 25\,{\lambda}^{2}+64\,m\lambda+72\,{m}^{2} \right) 
\cos \left( 2\,\theta \right) }{160\, \left( \lambda-2\,m \right) ^{2}
{\lambda}^{4}}}\right.\nonumber\\
&&\left.+{\frac {3\,\ln  \left( 1-\frac{2\,m}{\lambda} \right) }{8\,{m}^{3}}}+{\frac {240\,{\lambda}^{5}-720\,{\lambda}^{4}m+320\,{\lambda}^{3}{m}^{
2}+225\,{\lambda}^{2}{m}^{3}-96\,\lambda\,{m}^{4}+72\,{m}^{5}}{320\,{m
}^{2}{\lambda}^{4} \left( \lambda-2\,m \right) ^{2}}}
\right]a^4+{O}(a^6)\\
    \hat{r}&=&\lambda-\frac{(\lambda+m)\sin^2\theta}{2\lambda^2}a^2+\left[\frac{\sin^2\theta(5\cos2\theta+3))}{16\lambda^3}+\frac{m\sin^2\theta(7\cos2\theta+1)}{16\lambda^4}-\frac{m^2\sin^4\theta}{5\lambda^5}\right]a^4+{O}(a^6)\label{eq:rlambdatransfor}\\
    \hat{\theta}&=&\theta-{\frac {\sin \left( 2\,\theta \right) }{4\,{\lambda}^{2}
}}{a}^{2}+{\frac {\sin \left( 2\,\theta \right)  \left( 3\,\lambda\,\cos
 \left( 2\,\theta \right) +m\cos \left( 2\,\theta \right) -m \right)} {16\,{\lambda}^{5}}}a^{4}+{O}(a^6)\label{eq:thetatranfor}
\\
    \hat{\phi}&=&\phi+\left[\frac{1}{\lambda}+\frac{\ln(1-\frac{2m}{\lambda})}{2m}\right]a+\left[\frac{\ln(1-\frac{2m}{\lambda})}{4m}+\frac{m(2m+5\lambda)\cos2\theta}{8(\lambda-2m)\lambda^4}\right.\nonumber\\
    &&\left.-\frac{6m^4-m^3\lambda+8m^2\lambda^2+12m\lambda^3-12\lambda^4}{24m^2(\lambda-2m)\lambda^4}\right]a^3+{O}(a^5)
\end{eqnarray}
\end{widetext}

Finally, with these transformations in hand, we obtain the  same metric components in affine-null coordinates up to fourth order in $a$ as given in \eqref{eq:srot_metric_final} in the previous Section. 

\section{Localizing the  event horizon and ergosphere in affine-null cordinates}\label{sec:affine_null_kerr_trafo_hor}
In this Section we show that the affine-null coordinates for the slowly rotating Kerr metric  cover the ergosphere and the (past) event horizon $r_+$. 
In order to find them in a consistent way, they must be localized at $O(a^4)$. 
Recall that in BL coordinates the Kerr metric has the external ergosphere placed at 
\begin{equation}\label{eq:blergos}
\begin{split}
\hat{r}_{erg}=&m+\sqrt{m^2-a^2\cos^2\hat{\theta}}\\=&2m-\frac{a^2\cos^2\hat{\theta}}{2m}
-\frac{a^4\cos^4\hat{\theta}}{8m^3}+{O}(a^6),
\end{split}
\end{equation} 
and the event horizon at 
\begin{equation}\label{eq:revento}
r_+=m+\sqrt{m^2-a^2}=2m-\frac{a^2}{2m}-\frac{a^4}{8m^3}+{O}(a^6).
\end{equation}
The boundary of the external ergosphere  is obtained by looking for the timelike surface $\Gamma$ where the stationary Killing vector field  $\partial_u$ becomes a null vector field that is where 
\begin{equation}\label{eq:ergos}
g_{uu}|_\Gamma=0.
\end{equation}
Taking into account the expression for $g_{uu}$ as found in the first equation of \eqref{eq:srot_metric_final}, the ergosphere will be located at a given $\lambda=\lambda_{erg}(\theta)$, with 
\begin{equation}\label{eq:explambdaerg}
    \lambda_{erg}(\theta)=\sum^2_{i=0}\lambda_{erg[2i]}(\theta)a^{2i}+O(a^{6}).
\end{equation}
where the even expansion is a consequence of the symmetry assumption of Sec.~\ref{sec:aff_null}.
Inserting \eqref{eq:explambdaerg} into \eqref{eq:ergos}, and after re-expanding
in powers of $a$ we find
\begin{equation}\label{eq:finallambdaerg}
\begin{split}
\lambda_{erg}(\theta)=&2\,m -\frac {  \left( 7\cos
 ^2\theta-3 \right) }{8m}a^2\\
 &-\frac { \left( 51 \cos^4\theta-2\cos^2\theta+31
 \right) }{640\,{m}^3} a^4+O(a^6),
\end{split}
\end{equation}
which gives the location of the (external) ergosphere in affine-null coordinates.
By replacing \eqref{eq:finallambdaerg} into \eqref{eq:rlambdatransfor} (using  the inverse of \eqref{eq:thetatranfor} to relate $\theta$ with $\hat\theta$) , and after a re-expansion in powers of $a$ it can be checked that the standard fourth order expression for the BL expression of the ergosphere as given by \eqref{eq:blergos} is recovered.

Similarly, for the (Killing) event horizon we search a null surface $\Sigma$ described in affine-null coordinates by $\Sigma(\lambda,\theta)=\lambda-\lambda_H(\theta)=0$. Hence, its normal vector  $N_a=\nabla_a\Sigma$ must satisfy $N^a N_a=0$ which implies the following differential equation for $\lambda_H(\theta)=0$,
\begin{equation}\label{eq:lambdah}
g^{ab}N_aN_b=W+2W^\theta\frac{\partial\lambda_H(\theta)}{\partial\theta}+\frac{h^{\theta\theta}}{R^2}\left(\frac{\partial\lambda_H(\theta)}{\partial\theta}\right)^2=0.
\end{equation}
Let us assume an expansion for $\lambda_H(\theta)$ similar to
\eqref{eq:explambdaerg}, i.e.
\begin{equation}\label{eq:explambdaH1}
    \lambda_{H}(\theta)=\sum^2_{i=0}\lambda_{H[2i]}(\theta)a^{2i}+O(a^{6});
\end{equation}
with $\lambda_{H[0]}=2m$  (the Schwarzschild value for the location of the horizon).
Introducing \eqref{eq:explambdaH1} into \eqref{eq:lambdah}; re-expanding again in powers of $a$, we find (omitting the $O(a^6)$ term)
\begin{widetext}
\begin{equation}
    \begin{split}
        0=&\left( \frac{\lambda_{H[2]}}{2m} + \frac{3\cos^2\theta + 1}{16m^2}\right)a^2\\
        &+\left[\frac{(\lambda_{H[2],\theta})^2}{4m^2} 
        - \frac{3\sin\theta\cos\theta }{8m^3}\lambda_{H[2],\theta}- \frac{\lambda_{H[2]}^2}{ 4m^2}
        - \frac{3(\cos^2\theta + 1)}{ 16m^3} \lambda_{H[2]}
        - \frac{127\cos^4\theta 
        - 320m^3 \lambda_{H[4]}
        - 84\cos^2\theta - 3}{ 640m^4}\right]a^4.     
    \end{split}
\end{equation}
\end{widetext}
So that solving for the coefficient $\lambda_{H[2]}$ and $\lambda_{H[4]}$ gives us
\begin{equation}
\begin{split}
    \lambda_H(\theta)=&2m-\frac{(1+3\cos^2\theta)}{8m}a^2\\
&+\frac {(29\, \cos^4\theta -78
 \cos^2\theta-31)}{640m^3}a^4+{O}(a^6),
    \end{split}
\end{equation}
which gives the location of the (past) event horizon in affine-null coordinates. By replacing into \eqref{eq:rlambdatransfor} and after a reexpansion in $a$  up to fourth order, the well known value \eqref{eq:revento} for the BL radial coordinate of the event horizon is recovered.  
At this location, the affine-null coordinate system is regular.
\section{Summary}\label{sec:conclusion}
We have derived high-order slow rotation approximation of the Kerr metric in affine-null coordinates. 
To achieve this aim a metric in affine-null coordinates was expanded off a spherically symmetric background metric that corresponds to a Schwarzschild metric in outgoing Eddington Finkelstein coordinates. 
This quasi-spherical expansion was done with respect to a general smallness parameter $\varepsilon$. 
Subject to stationarity and axial symmetry the perturbations did not depend on the $u$ and $\phi$ coordinate. 
Moreover, requiring even parity of the Komar integral of stationary (giving the mass of the system) and odd parity of the Komar integral of axial symmetry (giving the angular momentum of the system), we argued that on the one hand the metric functions $\gamma$, $R$, $W^\theta$ and $W$ have only even perturbations in $\varepsilon$ while on the other hand the metric fields $\delta$ and $W^\phi$ have only odd perturbations in $\varepsilon$. 
This fact significantly simplifies the integration  of the perturbation equations resulting form the quasi-spherical expansion of the Ricci tensor. 
In addition, we find that the integration of the perturbation equations follows an alternative hierarchical structure.
Meaning with the spherically symmetric background solution at hand, the linear perturbations only involve the functions $\delta$ and $W^\phi$ and its integration provides (after application of the boundary condition of an asymptotic inertial observer) one free integration constant $B$. 
At next order, the quadratic perturbations turn out to be a linear combination of the derivatives of functions $\gamma$, $R$, $W^\theta$ and $W$ together with nonlinear terms containing the integration constants $A$ of the background model and the free integration constant $B$ of the linear perturbation. Their integration also yields a free integration constant, $C$. Following up the next order, there  only  differential equations involving  the cubic perturbations of $\delta$ and $W^\phi$ as well as the integration constants $A$, $B$ and $C$ characterizing the lower order perturbations. This alternating scheme between the perturbations of $(\delta,W^\phi)$ and those of $(\gamma, R, W^\theta, W)$ continues up to any order and is in fact a result of the symmetry assumptions. 
A common feature in solving for the even and odd-parity modes of $\varepsilon$, is that at any order there is a fourth order master equation for either the perturbation in $\gamma$ or the perturbation in $\delta$. 
With the solution of this master equation, the remaining perturbations can be solve by mere integration. 
After having obtained the perturbed solution and calculation of the Komar mass and Komar angular momentum, the  arising free integration constants $A, B, C, ...$ can expressed by the Komar mass and Komar angular momentum or by mass and specific angular momentum. Hence, the solution depends only on two free physical parameters. 
The fact that the derived solution is depending only on two parameters goes along with the  black holes uniqueness theorems stating that any stationary and axially symmetric  vacuum solution of  Einstein equations is uniquely determined by two parameters characterising the mass and angular momentum of the black hole. Here we have required the solutions of occurring master equations to be finite (see discussion around \eqref{gen_Sol_psi_LP}) at the affine parameter value $\lambda=A/2$. This is the position of the past event horizon of the nonrotating solution and similar to Carter's requirement of having an non-degenerate horizon\cite{Carter:1971zc,Robinson:1975bv,heusler1996black}. 
Since the Komar angular momentum is $O(\varepsilon)$, it turns out that the formal expansion parameter $\varepsilon$ relates to  the specific angular momentum and the previously made quasi-spherical approximation is in fact a slow rotation approximation, like those of Hartle and Thorne \cite{1967ApJ...150.1005H,1968ApJ...153..807H}.
By successively solving Einstein equations, we thus derived a slow rotation approximation of the Kerr-metric up to fourth order in the specific angular momentum. 
This solution is further verified for correctness using a 'standard' approach by obtaining a different representation of a given metric in another coordinate chart via a coordinate transformation. 
The slowly rotating Kerr metric presented here also obeys the peeling property, which can be seen considering the Weyl scalars in \eqref{eq:Weylscalars} 
\begin{subequations}\label{eq:Weylscalars}
\begin{align}
    \Psi_0 &=  \left(\frac{3ma^2}{\lambda^5}+i\frac{15 m a^3 }{\lambda^6}\cos\theta\right)\sin^2\theta + O(\lambda^{-7})\\
    \Psi_1 &=  i \frac{3\sqrt{2}ma}{2\lambda^4}\sin\theta+O(\lambda^{-5})\\
    \Psi_2 &=  -\frac{m}{\lambda^3}
    -i\frac{3ma\cos\theta}{\lambda^4}+O(\lambda^{-5})\\
    \Psi_3 &= -i\frac{3\sqrt{2} ma}{4\lambda^4}\sin\theta+O(\lambda^{-5}) \\
    \Psi_4 &= 
    %\red{\frac{3\sqrt{2} ma^2}{4\lambda^5}\sin^2\theta}
    \frac{3  ma^2}{4\lambda^5}\sin^2\theta+O(\lambda^{-6})
\end{align}
\end{subequations}
We can see in \eqref{eq:Weylscalars} that $\Psi_4$ and $\Psi_3$ have a stronger fall-off as required by the peeling property stating that $\Psi_n\sim \lambda^{5-n}$ at large radii. This stronger fall-off is because of the requirement of stationarity, the metric is not depending on $u$  and the multipole structure of the solution \cite{Janis:1965tx}. To recall, for example, in a most general spacetime satifying the peeling property $\psi_4\sim (\partial^2_u \sigma)/\lambda$ where $
\sigma$ is  the gravitational strain   (e.g. the gravitational wave) as measured by an asymptotic observer.

Moreover it is easily checked that the (only) conserved Newman Penrose constant \cite{2006PhRvD..73h4023B} vanishes \cite{2007PhRvD..75d4003B}.

What is is interesting to remark is that up to the considered order of approximation of our work and those of \cite{2007PhRvD..75d4003B}, the small $a$ expansion and the large $\lambda$ expansion coincide. 
It would be interesting to see up until which order this is the case. 
Such analysis might give insight on the validity and universality of general small parameter expansions of the Kerr spacetime in relation to null coordinates. It may also give insight if a closed form solution of the Kerr metric  with a surface forming null coordinate can be obtained at all. 
The method presented here offers the possibility to calculate any type of approximate rotating null-metric solution that is stationary, axially symmetric and has a known spherically symmetric background, like e.g. those  to describe compact matter systems or with a cosmological constant. 
Indeed, the study presented  here (solving the characteristic equations in this affine-null, metric formulation for vacuum spacetimes) is the natural starting point for further studying matter system under the given  symmetry assumptions.  
Some of such questions we are currently investigating. 

\section*{Acknowledgements}
 The authors thanks J. Winicour, L. Lehner, N. Stergioulas, E. M\"uller and G. Dotti for discussions at 
(early) stages of the project. T.M acknowledges financial support from the  FONDECYT de iniciaci\'on 2019 (Project No. 11190854) of the "Agencia Nacional de Investigaci\'on y Desarrollo" in Chile. E.G  gratefully acknowledges the hospitality extended to him during his stay at the Facultad de Ingeniería, Universidad Diego Portales and the financial support from CONICET and SeCyT-UNC. {We also appreciate a communication with Berend Schneider pointing out an error in the preprint version of the article.}
\appendix
\section{Useful Relations between Legendre Polynomials}
\label{sec:app_LP}
For completeness, we list some properties of the associated Legendre differential equations and relations between the Legendre polynomials.
%The Legendre differential equation for the Legendre polynomials $P_\ell(y)$ is 
%\begin{equation}\label{LP_ODE}
%    \frac{d}{dy}\left[(1-y^2)\frac{dP_\ell}{dy}\right] + \ell(\ell+1)P_\ell=0
%\end{equation}
%where the Legendre Polynomials $P_\ell(y)$ are defined% \cite{somebook}
%\begin{equation}\label{def_LP_ell}
%    P_\ell(y) = \frac{1}{2^\ell \ell!}\frac{d^\ell}{dy^\ell}(y^2-1)^\ell
%\end{equation}
The associated Legendre differential equation is 
\begin{equation}\label{aLP_ODE}
    \frac{d}{dy}\left[(1-y^2)\frac{dP^{\hat{m}}_\ell}{dy}\right] + \left[\ell(\ell+1)-\frac{\hat {m}^2}{1-y^2}\right]P^{\hat m}_\ell=0
\end{equation}
where $P^{\hat {m}}_\ell(y)$ are the associated Legendre polynomials, defined via 
\begin{equation}\label{def_LP_ell_m}
\begin{split}
    P^{\hat {m}}_\ell(y) 
     =&\frac{(-)^{\hat{m}}}{2^\ell \ell!}
    (1-y^2)^{\hat{m}/2}
     \frac{d^{\ell+\hat m}}{dy^{\ell+\hat m}}(y^2-1)^\ell
\end{split}
\end{equation}
In particular if $\hat {m}=0$ we have $P^0_\ell(y) = P_\ell(y)$, which are the well known Legendre polynomials.
From these definitions, some useful identities can be derived
\begin{equation}\label{rel1_aLP}
    \frac{d}{dy}\left[(1-y^2)P^2_\ell(y)\right] = [\ell(\ell+1)-2](1-y^2)^{1/2}P^1_\ell(y)
\end{equation}
\begin{eqnarray}\label{rel2_aLP}
    \frac{\frac{d}{dy}\Big[(1-y^2)^2\frac{dP^2_\ell}{dy}\Big]}{1-y^2} -2P^2_\ell
   &=&\ell(\ell+1)(\ell+2)(\ell-1)P_\ell(y)\nonumber\\
   \\
   P^1_\ell
   &=&
  -(1-y^2)^{1/2}\frac{dP^0_\ell}{dy}\label{rel3a_aLP}\\
  P^2_\ell
   &=&
  (1-y^2) \frac{d^2P^0_\ell}{dy^2}\label{rel3b_aLP}
 \end{eqnarray}
\begin{equation}\label{rel4_aLP}
    \frac{d}{dy}\left[(1-y^2)^2\frac{d}{dy}\frac{P^1_\ell}{(1-y^2)^{1/2}}\right] = [2-\ell(\ell+1)](1-y^2)^{1/2} P^1_\ell
\end{equation}
\begin{equation}\label{rel5_aLP}
     \frac{d}{dy}(1-y^2)^{1/2}P^1_\ell
     = \ell(\ell+1) P^0_\ell
\end{equation}

\section{Komar charges}\label{sec:app_komar_charge}
Depending on the Killing vector $X^a\in\{\partial_u, \partial_\phi\}$, we take the Komar charges to be 
\begin{equation}
K_{X} = -\frac{k_X}{8\pi}\oint \nabla^{[a}X^{b]}d\Sigma_{ab}
\end{equation}
with $k_{X} = 1, -1/2$ for a timelike ( e.g. $\ \partial_u$) or rotational Killing vector (e.g.   $\partial_\phi$), respectively.
Consider the  general  null metric with the nonzero
contravariant components $g^{01}$, $g^{11}$, $g^{1A}$ and $g^{AB}$.
The corresponding line element is 
\begin{equation}
    \begin{split}
        g_{ab}dx^adx^b &=
        ( g^{11}+g_{AB}g^{1A}g^{1B} )\left(\frac{dx^0}{g^{01}}\right)^2 + 2\left(\frac{dx^0}{g^{01}}\right)dx^1 \\
        &
        -2g_{AB}g^{1A}dx^B\left(\frac{dx^0}{g^{01}}\right)
        +g_{AB}dx^Adx^B
    \end{split}
\end{equation} 
where $g_{AC}g^{CB}=\delta_A^B$.
We define the null vectors $l$ and $n$ which obey $l^an_a  +1=l^al_a = n_an^a = 0$ as
\begin{equation}
     l=l^a\partial_a = -g^{01}\partial_1\;\;,\;\;
    n=n^a\partial_a=\partial_0  +\frac{1}{2}\frac{g^{11}}{g^{01} }\partial_1
    +\frac{g^{1A}}{g^{01}}\partial_A
\end{equation}
We note that $l$ points into the future. The associated  
covariant components are 
\begin{equation}
   % l = -dx^0 = -g^{01}\partial_1
   l_adx^a = -dx^0
   \;\;,\;\;
   n_adx^a  
    = -\frac{1}{2}\frac{g^{11}}{(g^{01})^2}dx^0 
    + \frac{dx^1}{g^{01}} 
\end{equation}
respectively.
The surface element $d\Sigma_{ab}$ follows as
\begin{equation}
d\Sigma_{ab} =2 l_{[a}n_{b]}\sqrt{\det(g_{AB})}dx^2dx^3
\end{equation}
with $x^A=(x^2, x^3)$ being any angular coordinates for the units sphere.
Setting $g_{AB} = R^2h_{AB}$ with $h_{AB}$ having the determinant of the unit sphere metric $q_{AB}$,  $q(x^C):=\det(h_{AB}) $. The corresponding volume element is defined as $d^2q:=\sqrt{q}dx^2dx^3$   and we have
$\oint  d^2q = 4\pi.$
Hence, 
$$d\Sigma_{ab}  
=2 l_{[a}n_{b]} R^2 d^2q\;\;.
$$
This allows us to write the Komar integal as
\begin{equation}
    K(X) = -\frac{k_X}{8\pi}\oint (2l^a n^{b} \partial_{[a}X_{b]}) R^2 d^2q\;\;,\;\;
\end{equation}
Since
\begin{eqnarray}
   2 l^a n^{b} \partial_{[a}X_{b]}
   &=&
   2l^{[a} n^{b]}  X_{b,a}
\\&=&
   (l^{a} n^{b} - l^bn^a)  X_{b,a}
\\&=&
   l^1 (n^{b}X_{b,1})
  - l^1 (n^{b}  X_{1,b})
   \\&=&
    -g^{01}[ (n^{b}X_{b,1})
   -(n^{b}  X_{1,b})]\;\;,
\end{eqnarray}
we have
\begin{equation}
    K(X) = 
   % \frac{k_X}{8\pi}\oint  I(X) g^{01} R^2 d^2q ,
    \frac{k_X}{8\pi}\oint \Big [ (n^{b}X_{b,1})
   -(n^{b}  X_{1,b})\Big]  g^{01} R^2 d^2q ,
\end{equation}
Taking the Killing vector to be $X = X^a\partial_a$
and specification to an affine null metric
\begin{equation}
\begin{split}
    g^{01}& = \epsilon\;,\;
g^{1A} = \epsilon W^A\;,\;
g^{11} = W\;,\;\\
g_{0A}& = -R^2h_{AB}W^B\;,\;
g_{AB} = R^2h_{AB}
\end{split}    
\end{equation}{}
and $\epsilon^2=1$ gives us
\begin{align}
    2 l^a n^{b} \partial_{[a}X_{b]}
    &=
      -\Big[ 
        W_{,1}
      -R^2h_{AB}W^AW^B_{,1}
      \Big]X^0
    \nonumber\\&
    +R^2\Big(h_{AB}W^B_{,1}X^A 
          -2h_{AB}W^B X^A_{,1}
          \Big)
    \nonumber\\&
   +\epsilon( X^1_{,1} -  X^0_{,0})
    -WX^0_{,1}
    -W^A X^0_{,A}
\end{align}

Assuming the timelike Killing vector $X =  \partial_0$
gives us
\begin{eqnarray*}
    2 l^a n^{b} \partial_{[a}X_{b]}
    &=&
      -\Big[ 
        W_{,1}
      -R^2h_{AB}W^AW^B_{,1}.
      \Big]
\end{eqnarray*}
Thus for the above form of the Killing vector we have the related Komar charge using $k_X = 1$
\begin{equation}
K(\partial_0) = \frac{1 }{8\pi}\oint \Big(  -\epsilon\Big[ 
        W_{,1}
      -R^2h_{AB}W^AW^B_{,1}
      \Big] \Big) R^2 d^2q . 
\end{equation}{}
%\subsection{Axial Killing vector}
With the rotational Killing $X = \partial_3$, we have
\begin{eqnarray*}
    2 l^a n^{b} \partial_{[a}X_{b]}
    &=&
    %R^2 h_{3B}W^B_{,1}\xi^3 
    %=
    R^2 h_{3B}W^B_{,1}
\end{eqnarray*}
so that the Komar charge is with $k_X = -\frac{1}{2}$
\begin{equation}
K(\partial_3)
=
-\frac{\epsilon}{16\pi}\oint \Big(R^4 h_{3B}W^B_{,1}\Big) d^2q. 
\end{equation}

%\input{answer_ref}

%\bibliography{ref}

\begin{thebibliography}{45}
\expandafter\ifx\csname natexlab\endcsname\relax\def\natexlab#1{#1}\fi
\expandafter\ifx\csname bibnamefont\endcsname\relax
  \def\bibnamefont#1{#1}\fi
\expandafter\ifx\csname bibfnamefont\endcsname\relax
  \def\bibfnamefont#1{#1}\fi
\expandafter\ifx\csname citenamefont\endcsname\relax
  \def\citenamefont#1{#1}\fi
\expandafter\ifx\csname url\endcsname\relax
  \def\url#1{\texttt{#1}}\fi
\expandafter\ifx\csname urlprefix\endcsname\relax\def\urlprefix{URL }\fi
\providecommand{\bibinfo}[2]{#2}
\providecommand{\eprint}[2][]{\url{#2}}

\bibitem[{\citenamefont{{Bondi}}(1960)}]{1960Natur.186..535B}
\bibinfo{author}{\bibfnamefont{H.}~\bibnamefont{{Bondi}}},
  \bibinfo{journal}{\nat} \textbf{\bibinfo{volume}{186}}, \bibinfo{pages}{535}
  (\bibinfo{year}{1960}).

\bibitem[{\citenamefont{{Bondi} et~al.}(1962)\citenamefont{{Bondi}, {van der
  Burg}, and {Metzner}}}]{1962RSPSA.269...21B}
\bibinfo{author}{\bibfnamefont{H.}~\bibnamefont{{Bondi}}},
  \bibinfo{author}{\bibfnamefont{M.~G.~J.} \bibnamefont{{van der Burg}}},
  \bibnamefont{and} \bibinfo{author}{\bibfnamefont{A.~W.~K.}
  \bibnamefont{{Metzner}}}, \bibinfo{journal}{Proceedings of the Royal Society
  of London Series A} \textbf{\bibinfo{volume}{269}}, \bibinfo{pages}{21}
  (\bibinfo{year}{1962}).

\bibitem[{\citenamefont{{Sachs}}(1962)}]{1962RSPSA.270..103S}
\bibinfo{author}{\bibfnamefont{R.~K.} \bibnamefont{{Sachs}}},
  \bibinfo{journal}{Proceedings of the Royal Society of London Series A}
  \textbf{\bibinfo{volume}{270}}, \bibinfo{pages}{103} (\bibinfo{year}{1962}).

\bibitem[{\citenamefont{{Kerr}}(1963)}]{1963PhRvL..11..237K}
\bibinfo{author}{\bibfnamefont{R.~P.} \bibnamefont{{Kerr}}},
  \bibinfo{journal}{\prl} \textbf{\bibinfo{volume}{11}}, \bibinfo{pages}{237}
  (\bibinfo{year}{1963}).

\bibitem[{\citenamefont{{Newman} et~al.}(1965)\citenamefont{{Newman}, {Couch},
  {Chinnapared}, {Exton}, {Prakash}, and {Torrence}}}]{1965JMP.....6..918N}
\bibinfo{author}{\bibfnamefont{E.~T.} \bibnamefont{{Newman}}},
  \bibinfo{author}{\bibfnamefont{E.}~\bibnamefont{{Couch}}},
  \bibinfo{author}{\bibfnamefont{K.}~\bibnamefont{{Chinnapared}}},
  \bibinfo{author}{\bibfnamefont{A.}~\bibnamefont{{Exton}}},
  \bibinfo{author}{\bibfnamefont{A.}~\bibnamefont{{Prakash}}},
  \bibnamefont{and}
  \bibinfo{author}{\bibfnamefont{R.}~\bibnamefont{{Torrence}}},
  \bibinfo{journal}{Journal of Mathematical Physics}
  \textbf{\bibinfo{volume}{6}}, \bibinfo{pages}{918} (\bibinfo{year}{1965}).

\bibitem[{\citenamefont{{Jordan} et~al.}(2013)\citenamefont{{Jordan}, {Ehlers},
  and {Sachs}}}]{2013GReGr..45.2691J}
\bibinfo{author}{\bibfnamefont{P.}~\bibnamefont{{Jordan}}},
  \bibinfo{author}{\bibfnamefont{J.}~\bibnamefont{{Ehlers}}}, \bibnamefont{and}
  \bibinfo{author}{\bibfnamefont{R.~K.} \bibnamefont{{Sachs}}},
  \bibinfo{journal}{General Relativity and Gravitation}
  \textbf{\bibinfo{volume}{45}}, \bibinfo{pages}{2691} (\bibinfo{year}{2013}).

\bibitem[{\citenamefont{{G{\'o}mez} et~al.}(1998)\citenamefont{{G{\'o}mez},
  {Lehner}, {Marsa}, and {Winicour}}}]{1998PhRvD..57.4778G}
\bibinfo{author}{\bibfnamefont{R.}~\bibnamefont{{G{\'o}mez}}},
  \bibinfo{author}{\bibfnamefont{L.}~\bibnamefont{{Lehner}}},
  \bibinfo{author}{\bibfnamefont{R.~L.} \bibnamefont{{Marsa}}},
  \bibnamefont{and}
  \bibinfo{author}{\bibfnamefont{J.}~\bibnamefont{{Winicour}}},
  \bibinfo{journal}{\prd} \textbf{\bibinfo{volume}{57}}, \bibinfo{pages}{4778}
  (\bibinfo{year}{1998}), \eprint{gr-qc/9710138}.

\bibitem[{\citenamefont{{Winicour}}(2012)}]{2012LRR....15....2W}
\bibinfo{author}{\bibfnamefont{J.}~\bibnamefont{{Winicour}}},
  \bibinfo{journal}{Living Reviews in Relativity}
  \textbf{\bibinfo{volume}{15}}, \bibinfo{eid}{2} (\bibinfo{year}{2012}).

\bibitem[{\citenamefont{{M{\"a}dler} and
  {Winicour}}(2016{\natexlab{a}})}]{2016SchpJ..1133528M}
\bibinfo{author}{\bibfnamefont{T.}~\bibnamefont{{M{\"a}dler}}}
  \bibnamefont{and}
  \bibinfo{author}{\bibfnamefont{J.}~\bibnamefont{{Winicour}}},
  \bibinfo{journal}{Scholarpedia} \textbf{\bibinfo{volume}{11}},
  \bibinfo{pages}{33528} (\bibinfo{year}{2016}{\natexlab{a}}),
  \eprint{1609.01731}.

\bibitem[{\citenamefont{{Barnich} and
  {Troessaert}}(2010)}]{2010JHEP...05..062B}
\bibinfo{author}{\bibfnamefont{G.}~\bibnamefont{{Barnich}}} \bibnamefont{and}
  \bibinfo{author}{\bibfnamefont{C.}~\bibnamefont{{Troessaert}}},
  \bibinfo{journal}{Journal of High Energy Physics}
  \textbf{\bibinfo{volume}{2010}}, \bibinfo{eid}{62} (\bibinfo{year}{2010}),
  \eprint{1001.1541}.

\bibitem[{\citenamefont{{Pasterski} et~al.}(2016)\citenamefont{{Pasterski},
  {Strominger}, and {Zhiboedov}}}]{2016JHEP...12..053P}
\bibinfo{author}{\bibfnamefont{S.}~\bibnamefont{{Pasterski}}},
  \bibinfo{author}{\bibfnamefont{A.}~\bibnamefont{{Strominger}}},
  \bibnamefont{and}
  \bibinfo{author}{\bibfnamefont{A.}~\bibnamefont{{Zhiboedov}}},
  \bibinfo{journal}{Journal of High Energy Physics}
  \textbf{\bibinfo{volume}{2016}}, \bibinfo{eid}{53} (\bibinfo{year}{2016}),
  \eprint{1502.06120}.

\bibitem[{\citenamefont{{M{\"a}dler} and
  {Winicour}}(2016{\natexlab{b}})}]{2016CQGra..33q5006M}
\bibinfo{author}{\bibfnamefont{T.}~\bibnamefont{{M{\"a}dler}}}
  \bibnamefont{and}
  \bibinfo{author}{\bibfnamefont{J.}~\bibnamefont{{Winicour}}},
  \bibinfo{journal}{Classical and Quantum Gravity}
  \textbf{\bibinfo{volume}{33}}, \bibinfo{eid}{175006}
  (\bibinfo{year}{2016}{\natexlab{b}}), \eprint{1605.01273}.

\bibitem[{\citenamefont{{Nichols}}(2017)}]{2017PhRvD..95h4048N}
\bibinfo{author}{\bibfnamefont{D.~A.} \bibnamefont{{Nichols}}},
  \bibinfo{journal}{\prd} \textbf{\bibinfo{volume}{95}}, \bibinfo{eid}{084048}
  (\bibinfo{year}{2017}), \eprint{1702.03300}.

\bibitem[{\citenamefont{{M{\"a}dler} and
  {Winicour}}(2018)}]{2018CQGra..35c5009M}
\bibinfo{author}{\bibfnamefont{T.}~\bibnamefont{{M{\"a}dler}}}
  \bibnamefont{and}
  \bibinfo{author}{\bibfnamefont{J.}~\bibnamefont{{Winicour}}},
  \bibinfo{journal}{Classical and Quantum Gravity}
  \textbf{\bibinfo{volume}{35}}, \bibinfo{eid}{035009} (\bibinfo{year}{2018}),
  \eprint{1708.08774}.

\bibitem[{\citenamefont{{M{\"a}dler} and
  {Winicour}}(2019)}]{2019CQGra..36i5009M}
\bibinfo{author}{\bibfnamefont{T.}~\bibnamefont{{M{\"a}dler}}}
  \bibnamefont{and}
  \bibinfo{author}{\bibfnamefont{J.}~\bibnamefont{{Winicour}}},
  \bibinfo{journal}{Classical and Quantum Gravity}
  \textbf{\bibinfo{volume}{36}}, \bibinfo{eid}{095009} (\bibinfo{year}{2019}),
  \eprint{1811.04711}.

\bibitem[{\citenamefont{{Papadopoulos}}(2002)}]{2002PhRvD..65h4016P}
\bibinfo{author}{\bibfnamefont{P.}~\bibnamefont{{Papadopoulos}}},
  \bibinfo{journal}{\prd} \textbf{\bibinfo{volume}{65}}, \bibinfo{eid}{084016}
  (\bibinfo{year}{2002}), \eprint{gr-qc/0104024}.

\bibitem[{\citenamefont{{Bishop} et~al.}(1996)\citenamefont{{Bishop},
  {G{\'o}mez}, {Lehner}, and {Winicour}}}]{1996PhRvD..54.6153B}
\bibinfo{author}{\bibfnamefont{N.~T.} \bibnamefont{{Bishop}}},
  \bibinfo{author}{\bibfnamefont{R.}~\bibnamefont{{G{\'o}mez}}},
  \bibinfo{author}{\bibfnamefont{L.}~\bibnamefont{{Lehner}}}, \bibnamefont{and}
  \bibinfo{author}{\bibfnamefont{J.}~\bibnamefont{{Winicour}}},
  \bibinfo{journal}{\prd} \textbf{\bibinfo{volume}{54}}, \bibinfo{pages}{6153}
  (\bibinfo{year}{1996}).

\bibitem[{\citenamefont{{G{\'o}mez} et~al.}(1994)\citenamefont{{G{\'o}mez},
  {Papadopoulos}, and {Winicour}}}]{1994JMP....35.4184G}
\bibinfo{author}{\bibfnamefont{R.}~\bibnamefont{{G{\'o}mez}}},
  \bibinfo{author}{\bibfnamefont{P.}~\bibnamefont{{Papadopoulos}}},
  \bibnamefont{and}
  \bibinfo{author}{\bibfnamefont{J.}~\bibnamefont{{Winicour}}},
  \bibinfo{journal}{Journal of Mathematical Physics}
  \textbf{\bibinfo{volume}{35}}, \bibinfo{pages}{4184} (\bibinfo{year}{1994}),
  \eprint{gr-qc/0006081}.

\bibitem[{\citenamefont{{Siebel} et~al.}(2002)\citenamefont{{Siebel}, {Font},
  {M{\"u}ller}, and {Papadopoulos}}}]{2002PhRvD..65f4038S}
\bibinfo{author}{\bibfnamefont{F.}~\bibnamefont{{Siebel}}},
  \bibinfo{author}{\bibfnamefont{J.~A.} \bibnamefont{{Font}}},
  \bibinfo{author}{\bibfnamefont{E.}~\bibnamefont{{M{\"u}ller}}},
  \bibnamefont{and}
  \bibinfo{author}{\bibfnamefont{P.}~\bibnamefont{{Papadopoulos}}},
  \bibinfo{journal}{\prd} \textbf{\bibinfo{volume}{65}}, \bibinfo{eid}{064038}
  (\bibinfo{year}{2002}), \eprint{gr-qc/0111093}.

\bibitem[{\citenamefont{{M{\"a}dler} and
  {M{\"u}ller}}(2013)}]{2013CQGra..30e5019M}
\bibinfo{author}{\bibfnamefont{T.}~\bibnamefont{{M{\"a}dler}}}
  \bibnamefont{and}
  \bibinfo{author}{\bibfnamefont{E.}~\bibnamefont{{M{\"u}ller}}},
  \bibinfo{journal}{Classical and Quantum Gravity}
  \textbf{\bibinfo{volume}{30}}, \bibinfo{eid}{055019} (\bibinfo{year}{2013}),
  \eprint{1211.4980}.

\bibitem[{\citenamefont{{Winicour}}(2013)}]{2013PhRvD..87l4027W}
\bibinfo{author}{\bibfnamefont{J.}~\bibnamefont{{Winicour}}},
  \bibinfo{journal}{\prd} \textbf{\bibinfo{volume}{87}}, \bibinfo{eid}{124027}
  (\bibinfo{year}{2013}), \eprint{1303.6969}.

\bibitem[{\citenamefont{{M{\"a}dler}}(2019)}]{2019PhRvD..99j4048M}
\bibinfo{author}{\bibfnamefont{T.}~\bibnamefont{{M{\"a}dler}}},
  \bibinfo{journal}{\prd} \textbf{\bibinfo{volume}{99}}, \bibinfo{eid}{104048}
  (\bibinfo{year}{2019}), \eprint{1810.04743}.

\bibitem[{\citenamefont{{Gallo} et~al.}(2021)\citenamefont{{Gallo}, {Kozameh},
  {M{\"a}dler}, {Moreschi}, and {Perez}}}]{2021PhRvD.104h4048G}
\bibinfo{author}{\bibfnamefont{E.}~\bibnamefont{{Gallo}}},
  \bibinfo{author}{\bibfnamefont{C.}~\bibnamefont{{Kozameh}}},
  \bibinfo{author}{\bibfnamefont{T.}~\bibnamefont{{M{\"a}dler}}},
  \bibinfo{author}{\bibfnamefont{O.~M.} \bibnamefont{{Moreschi}}},
  \bibnamefont{and} \bibinfo{author}{\bibfnamefont{A.}~\bibnamefont{{Perez}}},
  \bibinfo{journal}{\prd} \textbf{\bibinfo{volume}{104}}, \bibinfo{eid}{084048}
  (\bibinfo{year}{2021}), \eprint{2107.10120}.

\bibitem[{\citenamefont{{Crespo} et~al.}(2019)\citenamefont{{Crespo}, {de
  Oliveira}, and {Winicour}}}]{2019PhRvD.100j4017C}
\bibinfo{author}{\bibfnamefont{J.~A.} \bibnamefont{{Crespo}}},
  \bibinfo{author}{\bibfnamefont{H.~P.} \bibnamefont{{de Oliveira}}},
  \bibnamefont{and}
  \bibinfo{author}{\bibfnamefont{J.}~\bibnamefont{{Winicour}}},
  \bibinfo{journal}{\prd} \textbf{\bibinfo{volume}{100}}, \bibinfo{eid}{104017}
  (\bibinfo{year}{2019}), \eprint{1910.03439}.

\bibitem[{\citenamefont{{Baake} and {M{\"a}dler}}(2023)}]{Baake2023}
\bibinfo{author}{\bibfnamefont{O.}~\bibnamefont{{Baake}}} \bibnamefont{and}
  \bibinfo{author}{\bibfnamefont{T.}~\bibnamefont{{M{\"a}dler}}},
  \bibinfo{journal}{{in prep. }}  (\bibinfo{year}{2023}).

\bibitem[{\citenamefont{{Bishop} and {Venter}}(2006)}]{2006PhRvD..73h4023B}
\bibinfo{author}{\bibfnamefont{N.~T.} \bibnamefont{{Bishop}}} \bibnamefont{and}
  \bibinfo{author}{\bibfnamefont{L.~R.} \bibnamefont{{Venter}}},
  \bibinfo{journal}{\prd} \textbf{\bibinfo{volume}{73}}, \bibinfo{eid}{084023}
  (\bibinfo{year}{2006}), \eprint{gr-qc/0506077}.

\bibitem[{\citenamefont{{Arga{\~n}araz} and
  {Moreschi}}(2021)}]{2021PhRvD.104b4049A}
\bibinfo{author}{\bibfnamefont{M.~A.} \bibnamefont{{Arga{\~n}araz}}}
  \bibnamefont{and} \bibinfo{author}{\bibfnamefont{O.~M.}
  \bibnamefont{{Moreschi}}}, \bibinfo{journal}{\prd}
  \textbf{\bibinfo{volume}{104}}, \bibinfo{eid}{024049} (\bibinfo{year}{2021}).

\bibitem[{\citenamefont{{Fletcher} and {Lun}}(2003)}]{2003CQGra..20.4153F}
\bibinfo{author}{\bibfnamefont{S.~J.} \bibnamefont{{Fletcher}}}
  \bibnamefont{and} \bibinfo{author}{\bibfnamefont{A.~W.~C.}
  \bibnamefont{{Lun}}}, \bibinfo{journal}{Classical and Quantum Gravity}
  \textbf{\bibinfo{volume}{20}}, \bibinfo{pages}{4153} (\bibinfo{year}{2003}).

\bibitem[{\citenamefont{{Jahanur Hoque} and
  {Virmani}}(2021)}]{2021arXiv210801098J}
\bibinfo{author}{\bibfnamefont{S.}~\bibnamefont{{Jahanur Hoque}}}
  \bibnamefont{and}
  \bibinfo{author}{\bibfnamefont{A.}~\bibnamefont{{Virmani}}},
  \bibinfo{journal}{arXiv e-prints} \bibinfo{eid}{arXiv:2108.01098}
  (\bibinfo{year}{2021}), \eprint{2108.01098}.

\bibitem[{\citenamefont{Hayward}(2004)}]{Hayward:2004ih}
\bibinfo{author}{\bibfnamefont{S.~A.} \bibnamefont{Hayward}},
  \bibinfo{journal}{Phys. Rev. Lett.} \textbf{\bibinfo{volume}{92}},
  \bibinfo{pages}{191101} (\bibinfo{year}{2004}), \eprint{gr-qc/0401111}.

\bibitem[{\citenamefont{Gallo and Moreschi}(2014)}]{Gallo:2014jda}
\bibinfo{author}{\bibfnamefont{E.}~\bibnamefont{Gallo}} \bibnamefont{and}
  \bibinfo{author}{\bibfnamefont{O.~M.} \bibnamefont{Moreschi}},
  \bibinfo{journal}{Phys. Rev. D} \textbf{\bibinfo{volume}{89}},
  \bibinfo{pages}{084009} (\bibinfo{year}{2014}), \eprint{1404.2475}.

\bibitem[{\citenamefont{Arga\~naraz and Moreschi}(2022)}]{Arganaraz:2022mks}
\bibinfo{author}{\bibfnamefont{M.~A.} \bibnamefont{Arga\~naraz}}
  \bibnamefont{and} \bibinfo{author}{\bibfnamefont{O.~M.}
  \bibnamefont{Moreschi}}, \bibinfo{journal}{Phys. Rev. D}
  \textbf{\bibinfo{volume}{105}}, \bibinfo{pages}{084012}
  (\bibinfo{year}{2022}).

\bibitem[{\citenamefont{{Bai} et~al.}(2007)\citenamefont{{Bai}, {Cao}, {Gong},
  {Shang}, {Wu}, and {Lau}}}]{2007PhRvD..75d4003B}
\bibinfo{author}{\bibfnamefont{S.}~\bibnamefont{{Bai}}},
  \bibinfo{author}{\bibfnamefont{Z.}~\bibnamefont{{Cao}}},
  \bibinfo{author}{\bibfnamefont{X.}~\bibnamefont{{Gong}}},
  \bibinfo{author}{\bibfnamefont{Y.}~\bibnamefont{{Shang}}},
  \bibinfo{author}{\bibfnamefont{X.}~\bibnamefont{{Wu}}}, \bibnamefont{and}
  \bibinfo{author}{\bibfnamefont{Y.~K.} \bibnamefont{{Lau}}},
  \bibinfo{journal}{\prd} \textbf{\bibinfo{volume}{75}}, \bibinfo{eid}{044003}
  (\bibinfo{year}{2007}), \eprint{gr-qc/0701171}.

\bibitem[{\citenamefont{{Gong} et~al.}(2007)\citenamefont{{Gong}, {Shang},
  {Bai}, {Cao}, {Luo}, and {Lau}}}]{2007PhRvD..76j7501G}
\bibinfo{author}{\bibfnamefont{X.}~\bibnamefont{{Gong}}},
  \bibinfo{author}{\bibfnamefont{Y.}~\bibnamefont{{Shang}}},
  \bibinfo{author}{\bibfnamefont{S.}~\bibnamefont{{Bai}}},
  \bibinfo{author}{\bibfnamefont{Z.}~\bibnamefont{{Cao}}},
  \bibinfo{author}{\bibfnamefont{Z.}~\bibnamefont{{Luo}}}, \bibnamefont{and}
  \bibinfo{author}{\bibfnamefont{Y.~K.} \bibnamefont{{Lau}}},
  \bibinfo{journal}{\prd} \textbf{\bibinfo{volume}{76}}, \bibinfo{eid}{107501}
  (\bibinfo{year}{2007}).

\bibitem[{\citenamefont{{Hartle}}(1967)}]{1967ApJ...150.1005H}
\bibinfo{author}{\bibfnamefont{J.~B.} \bibnamefont{{Hartle}}},
  \bibinfo{journal}{\apj} \textbf{\bibinfo{volume}{150}}, \bibinfo{pages}{1005}
  (\bibinfo{year}{1967}).

\bibitem[{\citenamefont{Tafel}(2022)}]{Tafel:2021kza}
\bibinfo{author}{\bibfnamefont{J.}~\bibnamefont{Tafel}},
  \bibinfo{journal}{Class. Quant. Grav.} \textbf{\bibinfo{volume}{39}},
  \bibinfo{pages}{115013} (\bibinfo{year}{2022}), \eprint{2105.09372}.

\bibitem[{\citenamefont{{G{\'o}mez} et~al.}(2001)\citenamefont{{G{\'o}mez},
  {Husa}, and {Winicour}}}]{2001PhRvD..64b4010G}
\bibinfo{author}{\bibfnamefont{R.}~\bibnamefont{{G{\'o}mez}}},
  \bibinfo{author}{\bibfnamefont{S.}~\bibnamefont{{Husa}}}, \bibnamefont{and}
  \bibinfo{author}{\bibfnamefont{J.}~\bibnamefont{{Winicour}}},
  \bibinfo{journal}{\prd} \textbf{\bibinfo{volume}{64}}, \bibinfo{eid}{024010}
  (\bibinfo{year}{2001}), \eprint{gr-qc/0009092}.

\bibitem[{\citenamefont{{van der Burg}}(1966)}]{1966RSPSA.294..112V}
\bibinfo{author}{\bibfnamefont{M.~G.~J.} \bibnamefont{{van der Burg}}},
  \bibinfo{journal}{Proceedings of the Royal Society of London Series A}
  \textbf{\bibinfo{volume}{294}}, \bibinfo{pages}{112} (\bibinfo{year}{1966}).

\bibitem[{\citenamefont{{M{\"a}dler}}(2013)}]{2013PhRvD..87j4016M}
\bibinfo{author}{\bibfnamefont{T.}~\bibnamefont{{M{\"a}dler}}},
  \bibinfo{journal}{\prd} \textbf{\bibinfo{volume}{87}}, \bibinfo{eid}{104016}
  (\bibinfo{year}{2013}), \eprint{1212.3316}.

\bibitem[{\citenamefont{{Dozmorov}}(1975)}]{1975Fiz....18...95D}
\bibinfo{author}{\bibfnamefont{I.~M.} \bibnamefont{{Dozmorov}}},
  \bibinfo{journal}{Fizika} \textbf{\bibinfo{volume}{18}}, \bibinfo{pages}{95}
  (\bibinfo{year}{1975}).

\bibitem[{\citenamefont{Carter}(1971)}]{Carter:1971zc}
\bibinfo{author}{\bibfnamefont{B.}~\bibnamefont{Carter}},
  \bibinfo{journal}{Phys. Rev. Lett.} \textbf{\bibinfo{volume}{26}},
  \bibinfo{pages}{331} (\bibinfo{year}{1971}).

\bibitem[{\citenamefont{Robinson}(1975)}]{Robinson:1975bv}
\bibinfo{author}{\bibfnamefont{D.~C.} \bibnamefont{Robinson}},
  \bibinfo{journal}{Phys. Rev. Lett.} \textbf{\bibinfo{volume}{34}},
  \bibinfo{pages}{905} (\bibinfo{year}{1975}).

\bibitem[{\citenamefont{Heusler et~al.}(1996)\citenamefont{Heusler, Goddard,
  and Yeomans}}]{heusler1996black}
\bibinfo{author}{\bibfnamefont{M.}~\bibnamefont{Heusler}},
  \bibinfo{author}{\bibfnamefont{P.}~\bibnamefont{Goddard}}, \bibnamefont{and}
  \bibinfo{author}{\bibfnamefont{J.}~\bibnamefont{Yeomans}},
  \emph{\bibinfo{title}{Black Hole Uniqueness Theorems}}, Cambridge Lecture
  Notes in Physics (\bibinfo{publisher}{Cambridge University Press},
  \bibinfo{year}{1996}), ISBN \bibinfo{isbn}{9780521567350},
  \urlprefix\url{https://books.google.com.ar/books?id=H4eXl9QODoAC}.

\bibitem[{\citenamefont{{Hartle} and {Thorne}}(1968)}]{1968ApJ...153..807H}
\bibinfo{author}{\bibfnamefont{J.~B.} \bibnamefont{{Hartle}}} \bibnamefont{and}
  \bibinfo{author}{\bibfnamefont{K.~S.} \bibnamefont{{Thorne}}},
  \bibinfo{journal}{\apj} \textbf{\bibinfo{volume}{153}}, \bibinfo{pages}{807}
  (\bibinfo{year}{1968}).

\bibitem[{\citenamefont{Janis and Newman}(1965)}]{Janis:1965tx}
\bibinfo{author}{\bibfnamefont{A.~I.} \bibnamefont{Janis}} \bibnamefont{and}
  \bibinfo{author}{\bibfnamefont{E.~T.} \bibnamefont{Newman}},
  \bibinfo{journal}{J. Math. Phys.} \textbf{\bibinfo{volume}{6}},
  \bibinfo{pages}{902} (\bibinfo{year}{1965}).

\end{thebibliography}

\end{document}